%% file: LastArxivVersion.tex
\documentclass[twocolumn,aps,prl,groupedaddress,superscriptaddress,amsmath,amssymb,noeprint]{revtex4-2}
\input{preamble_packages.tex}
\input{preamble_commands.tex}

\begin{document}
	
\title{Reservoir-Free Decoherence in Flying Qubits}

\author{Nicol\`{o} Piccione}
\email{nicolo’.piccione@units.it}
\affiliation{Université Grenoble Alpes, CNRS, Grenoble INP, Institut Néel, 38000 Grenoble, France}
\affiliation{Istituto Nazionale di Fisica Nucleare, Trieste Section, Via Valerio 2, 34127 Trieste, Italy}

\author{L\'{e}a Bresque}
\affiliation{Université Grenoble Alpes, CNRS, Grenoble INP, Institut Néel, 38000 Grenoble, France}
\affiliation{The Abdus Salam International Center for Theoretical Physics (ICTP), Strada Costiera 11, 34151 Trieste, Italy}

\author{Andrew N. Jordan}
\affiliation{Institute for Quantum Studies, Chapman University, Orange, CA, 92866, USA}
\affiliation{Department of Physics and Astronomy, University of Rochester, Rochester, NY 14627, USA}
\affiliation{The Kennedy Chair in Physics, Chapman University, Orange, CA 92866, USA}

\author{Robert S. Whitney}
\affiliation{Université Grenoble Alpes, CNRS,
LPMMC, Grenoble 38000, France}

\author{Alexia Auff\`{e}ves}
\affiliation{Université Grenoble Alpes, CNRS, Grenoble INP, Institut Néel, 38000 Grenoble, France}
\affiliation{MajuLab, CNRS–UCA-SU-NUS-NTU International Joint Research Laboratory}
\affiliation{Centre for Quantum Technologies, National University of Singapore, 117543 Singapore, Singapore}

\date{\today}

\begin{abstract}
An effective time-dependent Hamiltonian can be implemented by making a quantum system fly through an inhomogeneous potential, realizing, for example, a quantum gate on its internal degrees of freedom. However, flying systems have a spatial spread that will generically entangle the internal and spatial degrees of freedom, leading to decoherence in the internal state dynamics, even in the absence of any external reservoir. We provide formulas valid at all times for the dynamics, fidelity, and change of entropy for ballistic particles with small spatial spreads, quantified by $\Delta x$. This non-Markovian decoherence can be significant for ballistic flying qubits (scaling as $\Delta x^2$) but usually not for flying qubits carried by a moving potential well (scaling as $\Delta x^6$). 
We also discuss a method to completely counteract this decoherence for a ballistic qubit later measured.
\end{abstract}

\maketitle

\begin{figure}
\centering
\includegraphics[width=0.48\textwidth]{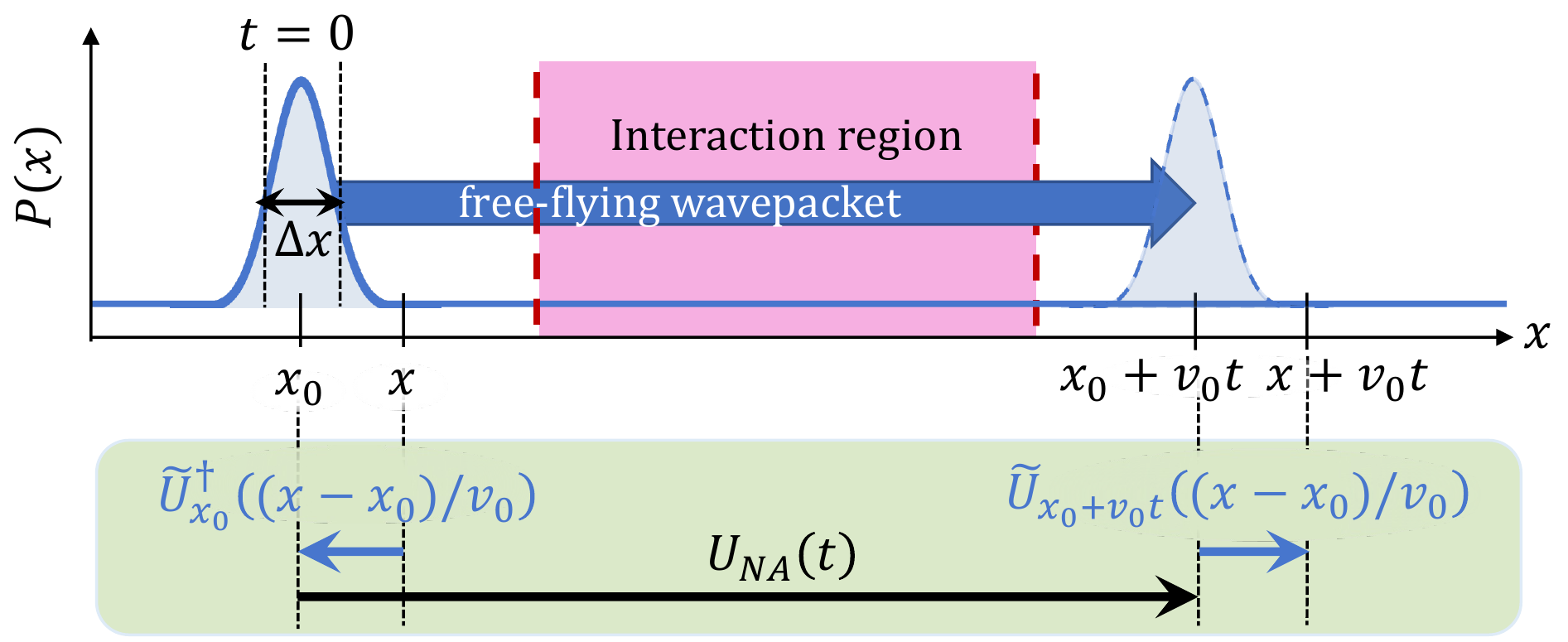}
\caption{A ballistic quantum system is moving at constant velocity to enter in and exit from an interaction region acting on its internal state. During the interaction the internal state associated with each position in the wavepacket evolves differently due to the different time spent in the interaction region and free evolution region. The green box shows a decomposition of the internal state dynamics associated with the point $x$. This decomposition is the one we used to obtain our results, see text for details.}
\label{fig:Illustration}
\end{figure}

Flying qubits, such as flying Rydberg atoms~\cite{Book_Haroche2006,Haroche2020Mar,chinese-collision,gate-gilles,arno-ghz,Najera-Santos2020Jul}, flying spin qubits \cite{Jadot2021Distant,Noiri2022Sep,Seidler2022Aug,Edlbauer2022semiconductor}, or flying electrons~\cite{Yamamoto2012Electrical,Bauerle2018Coherent,Takada2019Sound,Freise2020Trapping,Edlbauer2021Flight,Edlbauer2022semiconductor,Wang2023Jul,Henkel2022Jun}, have practical and fundamental significance. Practically, there is great hope to use the internal state of flying qubits to process and transport quantum information. This is a goal of recent experiments on flying electrons in solid state devices, with quantum information carried by the electron's spins \cite{Jadot2021Distant,Noiri2022Sep,Seidler2022Aug,Edlbauer2022semiconductor}, or its spatial distribution~\cite{Edlbauer2022semiconductor,Yamamoto2012Electrical}. Similar ideas have long been applied to flying Rydberg atoms~\cite{Book_Haroche2006,Haroche2020Mar,gate-gilles,chinese-collision,arno-ghz}. Fundamentally, they are the simplest examples of how time-dependent Hamiltonians emerge from time-independent ones, i.e., how non-autonomous dynamics emerge from autonomous ones~\cite{note_autonomous}
. This is used for measurement paradoxes~\cite{Aharonov1984Quantum,Aharonov1998Measurement,Gisin2018Quantum,rogers2022postselection, jordan2024quantum}, symmetries \cite{Whitney2008Mar}, quantum optics~\cite{Englert1991reflecting,Scully1996Induced,Larson2009Cavity,Mercurio2022Flying}, quantum collision models~\cite{Jacob2021Thermalization,Ciccarello2021Quantum,Jacob2022QuantumScattering,Tabanera2022Quantum}, and quantum thermodynamics~\cite{Book_Binder2018,DeChiara2018,Jacob2021Thermalization,Jacob2022QuantumScattering,Tabanera2022Quantum,Najera-Santos2020Jul}. As a system flies through a spatially varying potential, its internal state experiences a time-dependent Hamiltonian. However, this is \emph{only} true if the flying system is point-like~\cite{Jacob2022QuantumScattering}. Otherwise, its internal Degree of Freedom (DoF) also get entangled with its spatial DoF, causing decoherence of the internal DoF. This is a fundamental source of decoherence, intrinsic to the flying nature of the system. We remark that the internal DoF does not have to be a qubit (i.e., a two-level system); it can have arbitrary structure, the qubit being the paradigmatic case.

In this letter, we consider a quantum system that flies ballistically, i.e., at approximately constant velocity, with a small spatial spread, as drawn in Fig.~\ref{fig:Illustration}. We analyze its internal dynamics for arbitrary internal structure. The wavepacket's spatial spread causes noisy dynamics for the internal state, even in the absence of any external reservoirs. Thus, we refer to its effect as reservoir-free decoherence and we analytically characterize it.
We quantify this in terms of the internal state's fidelity (compared to an ideal point-like system), and its entropy change. Then, we apply our findings to flying qubits, and identify ways to reduce or completely nullify the reservoir-free decoherence. Finally, we estimate it for flying systems carried by a moving potential well, and find that it is much weaker than for ballistic flying systems.

\vskip 1mm
\emph{Ballistic system's dynamics.---}Consider a quantum system with arbitrary internal structure flying ballistically. We want its internal dynamics to correspond to a desired non-autonomous (i.e., time-dependent) dynamics given by the evolution operator $U_{\rm NA} (t)$, resulting from the time-dependent Hamiltonian $H_{\rm NA} (t)$. To do this, we let it fly through a one-dimensional potential so that the total, autonomous, Hamiltonian reads
\begin{equation}
\label{eq:StartingHamiltonian}
H = \frac{\hp^2}{2m} + H_0 + V(\hx),
\end{equation}
where $m$ is the system's mass, $\hx$ and $\hp$ are its position and momentum operators, $H_0$ is the $\hx$-independent part of the internal Hamiltonian and $V(\hx)$ the $\hx$-dependent part.

In the ideal case of a particle with classical spatial DoF, we can assign to it a definite position $x_{\rm cl}(t)$ and momentum at any time. Then, the interaction term acts on the internal DoF as $V(x_{\rm cl}(t))$. 
Assuming that the particle's momentum is constant, hence it moves at constant velocity $v_0$, its position is $x_{\rm cl}(t) = x_0 + v_0 t$, where $x_0$ is the particle's initial position. It follows that
the internal state evolves as governed by the Hamiltonian $H_{\rm NA} (t) \equiv H_0 + V(x_0+v_0t)$.
In the following, we consider what happens if the particle is not point-like, i.e., is described by a wavepacket of finite size initially centered at $x_0$. Then, $V(\hx)$ may also transfer energy between the particle's internal and spatial DoF, while entangling them.

\vskip 1mm
\emph{Approximations.---}Solving Eq.~(\ref{eq:StartingHamiltonian}) can be involved, even for wavepackets without internal DoF hitting simple barriers~\cite{Los2013Exact,Riahi2017Solving}. The regime we are interested in is defined by two approximations. Firstly, the kinetic energy changes induced by $V(\hx)$ are supposed negligible at all times with respect to the mean initial kinetic energy. Introducing $p_0$ the particle's initial average momentum and $\hq = \hp - p_0$, it yields $\ev{\hq} \ll p_0$ and $\ev{\hq^2} \ll p_0^2$. This leads to the so-called quantum clock dynamics \cite{Aharonov1984Quantum,Aharonov1998Measurement,Malabarba2015clock,Gisin2018Quantum,Soltan2021Conservation}, which corresponds to linearizing the kinetic energy in solid-state physics (used for flying qubits in \cite{Yamamoto2012Electrical}). Eq.~\eqref{eq:StartingHamiltonian} becomes
\begin{equation}
\label{eq:MasslessHamiltonian}
H \simeq v_0 \hq + H_0 +V(\hx),
\end{equation}
where we drop the constant $p_0^2/(2m)$, with no effect on the dynamics, and the approximation involves dropping $\hq^2/(2m)$. This means that the wavepacket propagates without dispersion and at a constant group velocity (further simply dubbed velocity) $v_0=p_0/m$.

Secondly, the particle should remain sufficiently localized at all times, such that its internal states typically accumulate a small phase difference over the entire wavepacket. Introducing $\Delta x$ the typical width of the wavepacket and $E_0$ the typical energy scale of the internal Hamiltonian $H_0+V(x)$, this condition means  
\begin{eqnarray}
\varepsilon \ \equiv\ \frac{\Delta x E_0}{\hbar v_0} \ \ll\ 1,
\end{eqnarray}
where the parameter $\varepsilon$ plays a major role in our calculations, as we show below. For Gaussian wavepackets, Heisenberg inequality is saturated ($\Delta x\Delta p = \hbar/2$), and $\varepsilon \ll 1$ can be written as $p_0 \Delta p/m \gg E_0$. It means that the spread of kinetic energy induced by the wavepacket localization largely overcomes the internal energy scale: hence, the different spatial states resulting from the evolution of different internal states remain almost indistinguishable - in other words, the spatial DoF carries a small amount of which path information on the internal DoF.

Starting from Eq.~\eqref{eq:MasslessHamiltonian}, the dynamics can be solved exactly, as shown in sec.~\ref{APPSec:ClockHamiltonianSolution} of the SM.
We consider the particle to be completely outside of the interaction region at $t=0$, as shown in fig.~\ref{fig:Illustration}. It makes then sense to consider spatial and internal DoF to be initially uncorrelated. The internal state at time $t$ is given by 
\begin{equation}
\label{eq:InternalState}
\rho_I (t) = \intmp A_0 (x,x) \tU_{x} (t) \rho_0 \tU_{x}^\dg (t) \dd{x}, 
\end{equation}
with $A_0 (x,x)$ the initial probability density of finding the particle at point $x$, $\rho_0$ the initial internal state, and 
\begin{equation}
\label{eq:UnitaryEvolutionPosition}
\tU_{x} (t)
=
\mathcal{T} \exp[- \frac{i}{\hbar} \int_{0}^{t} \dd{s} H_{\rm NA}\prt{s + \frac{x-x_0}{v_0}}],
\end{equation}
where $\mathcal{T}$ is the time-ordering operator. $\tU_{x} (t)$ is the evolution operator for the internal state associated with position $x$ in the wavepacket. 
Hence, different parts of the wavepacket (i.e., different $x$) have different dynamics, even though each part of the wavepacket goes through the same potential $V(\hx)$ during its flight.  This is the origin of the entanglement between the spatial and internal DoF, leading to the reservoir-free decoherence.

For a time $t_f$ such that the wavepacket has completely gone through the potential region, the resevoir-free decoherence can be intuitively explained as follows. For each initial position $x$ of the particle within the wavepacket, the internal state evolves according to the ideal dynamics one would have starting from $x$ instead of $x_0$. This dynamics splits into three parts: before, during, and after the interaction region. The interaction region acts in the same way for each starting position $x$, but the respective durations of the free evolution steps depend on $x$. If the dynamics in the interaction region does not commute with the free one, then each $x$ gives rise to a different total evolution, hence entangling the spatial and internal DoF. Otherwise, the evolution is the same for each starting point and there is no reservoir-free decoherence. This can happen, for example, if $\comm{V(\hx)}{H_0}=0$ or if $H_{\rm NA} (t)$ changes adiabatically.

\vskip 1mm
\emph{Approximate dynamics.---}We now solve the internal DoF dynamics in the regime of localized wavepacket defined by $\varepsilon \ll 1$. Let the flying particle's initial wavepacket be localized in space, centered at $x_0$ with a spread $\Delta x = [\ev{\hx^2}-\ev{\hx}^2]^{1/2}$, see Fig.~\ref{fig:Illustration}. The wavepacket moves from left to right at constant velocity $v_0$, so it is centered at $x_0 + v_0 t$ at time $t$. The part of the wavefunction initially at $x_0$ has internal evolution $U_{\rm NA} (t) \equiv \tU_{x_0} (t)$. For other initial $x$, we can decompose their evolution as shown in the green box of Fig.~\ref{fig:Illustration}; evolving from $x$ to $x_0$ using $\tU^\dg_{x_0} ((x-x_0)/v_0)$, from $x_0$ to $x_0 + v_0 t$ using $U_{\rm NA} (t)$, and from $x_0 + v_0 t$ to $x + v_0 t$ using $\tU_{x_0 + v_0 t} ((x-x_0)/v_0)$. Then, assuming $\varepsilon \ll 1$, we can expand $\tU^\dg_{x_0} ((x-x_0)/v_0)$ and $\tU_{x_0 + v_0 t} ((x-x_0)/v_0)$ by means of Taylor expansions (see secs.~\ref{APPSec:LocalizedWavePackets} and \ref{APPSec:Reducedstate} of SM). 
This gives simple expressions for the dynamics and quantities of interest, in the regime in which the internal state ideal dynamics is only weakly perturbed by the spatial spread of the wavepacket.

In this regime, the internal state's reduced density matrix at time $t$, after tracing over the spatial wavefunction (see sec.~\ref{APPSec:Reducedstate} of the SM), is 
\begin{equation}
\label{eq:ApproximateDensityMatrixIDoFExplicit}
\rC_I (t) \simeq
\rho_{\rm NA} (t) + \varepsilon^2 \mcC (t),
\end{equation}
where $\rho_{\rm NA} (t) = U_{\rm NA}(t) \rho_0 U_{\rm NA}^\dg (t)$ is the non-autonomous ideal dynamics (that of the wavepacket's center), and 
\begin{multline}
\label{eq:CorrectionTerm}
\mcC(t) = \bigl\{ \comm{H_{\rm NA}}{U \comm{H_0}{\rho_0} U^\dg} -
\frac{i \hbar}{2}\comm{\partial_t H_{\rm NA}}{U \rho_0 U^\dg}
\\
+U D_{H_0} (\rho_0) U^\dg + D_{H_{\rm NA}} \prt{U \rho_0 U^\dg} \bigr\}/ E_0^{2},
\end{multline}
is the correction term, with $D_{X} (\rho) = X \rho X^\dg - (1/2)\acomm{X^\dg X}{\rho}$, and $U$ (resp. $H_{\rm NA}$) being shorthand for $U_{\rm NA}(t)$ (resp. $H_\text{NA}(t)$). 
Importantly, Eqs.~(\ref{eq:ApproximateDensityMatrixIDoFExplicit}-\ref{eq:CorrectionTerm}) reveal that the deviation
from ideal dynamics scales as $\Delta x^2$ but its form (encoded in $\mcC(t)$) is independent of $\Delta x$. Moreover, at a practical level, Eqs.~(\ref{eq:ApproximateDensityMatrixIDoFExplicit}-\ref{eq:CorrectionTerm}) are easy to solve: unlike Eq.~\eqref{eq:StartingHamiltonian}, they do not involve the large Hilbert space of the spatial DoF, and standard perturbation theory can be applied to find $U_{\rm NA} (t)$ (see, e.g., Ref.~\cite{Bresque2022energetique}). Below we exploit these analytic expressions to quantify the impact of the reservoir-free decoherence.

\vskip 1mm
\emph{Fidelity and entropy.---}We consider two ways of characterizing how close the internal dynamics are to ideal~\cite{Book_Breuer2002,Book_Nielsen2010}: (i) the fidelity between real and ideal internal state, and (ii) the von Neumann entropy change of the real internal state. In sec. \ref{APPSec:Entropy} of the SM, we derive both from Eqs.~(\ref{eq:ApproximateDensityMatrixIDoFExplicit},\ref{eq:CorrectionTerm}), using a method from Ref.~\cite{Grace2021Perturbation}. When the initial internal state is pure, we define the ideal evolution as $\dyad{\psi_{\rm NA} (t)}$. Then, the fidelity $F (t) \equiv \ev{\rC_I (t)}{\psi_{\rm NA} (t)} $, and the von Neumann entropy $S (t) \equiv - \Tr \{ \rC_I (t) \ln \rC_I (t) \} $ are
\begin{align}
\label{eq:FidelityApproximatedFormula}
F(t) &\simeq
1 - \varepsilon^2\Big|\ev{\mcC(t)}{\psi_{\rm NA} (t)}\Big|,
\\
\label{eq:EntropyPerturbation}
S(t) &\simeq \varepsilon^2\ 
{\rm Tr} \prtg{\mcC_\perp(t)- \mcC_\perp(t)\ln\left[\varepsilon^2\mcC_\perp(t)\right]},
\end{align}
where $\mcC_\perp(t) = \big(1-\dyad{\psi_{\rm NA} (t)}\big)\mcC(t) \big(1-\dyad{\psi_{\rm NA} (t)}\big)$ is the part of $\mcC(t)$ orthogonal to $|\psi_{\rm NA} (t)\rangle$. The case of a mixed initial internal state is discussed in sec. \ref{APPSec:Entropy} of the SM. 
Eqs.~(\ref{eq:ApproximateDensityMatrixIDoFExplicit}-\ref{eq:EntropyPerturbation}) are the main results of this work.

\begin{figure}
	\centering
	\includegraphics[width=0.48\textwidth]{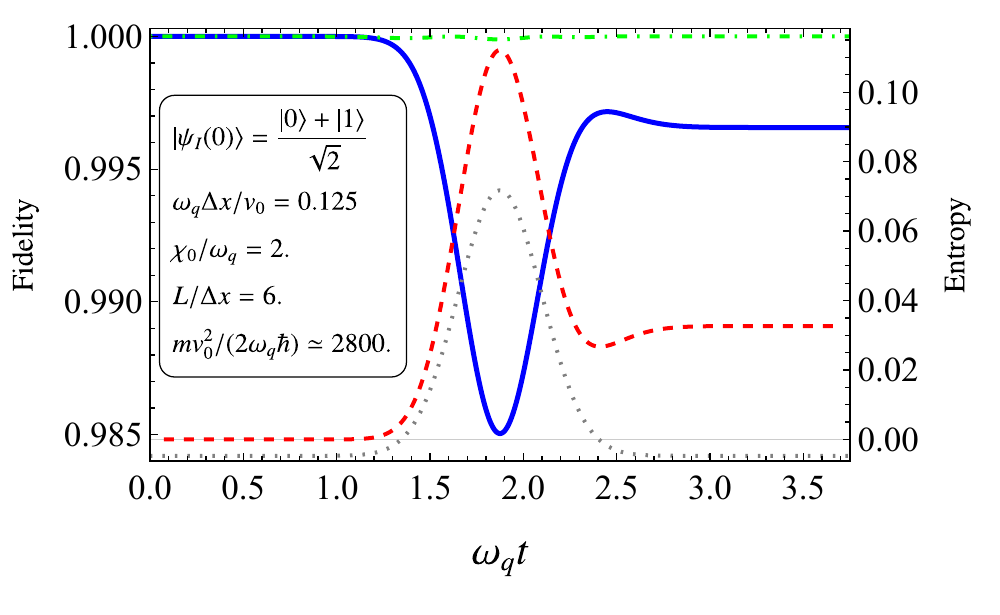}
	\caption{
 A qubit flying through an inhomogeneous potential. The qubit bare Hamiltonian is $H_0 = (1/2)\hbar \omega_q \sigma_z$ while the interaction term is $V(\hx)=\frac{1}{2}\hbar \chi_0\sigma_x\exp[-\pi x^2/L^2]$, sketched as the dotted gray curve. Since $H_{\rm NA}$ does not commute with itself at different times even the ideal dynamics are non-trivial, so we obtain both $\rho_{\rm NA}$ and $\rC_I$ numerically by adapting a method from Ref.~\cite{Book_Schmied2020Mathematica}. The initial state is a Gaussian wavepacket with spatial spread $\Delta x$ centered at $x_0$ and mean wavevector $k_0$, with internal state $\big(|0\rangle+|1\rangle\big)\big/\sqrt{2}$. The continuous blue line represents the fidelity in Eq.~(\ref{eq:FidelityApproximatedFormula}) as the qubit flies. The dotdashed green line is the fidelity of the approximate state Eq.~(\ref{eq:ApproximateDensityMatrixIDoFExplicit}) relative to an exact evolution of Eq.~(\ref{eq:StartingHamiltonian}); it is close to one, showing that the approximation is very good. Finally, the red dashed line represents the internal state's von Neumann entropy (in bits) in Eq.~\eqref{eq:EntropyPerturbation}.
 }
\label{fig:FidelityEntropyPlot}
\end{figure}

Eqs.~(\ref{eq:ApproximateDensityMatrixIDoFExplicit}-\ref{eq:EntropyPerturbation}) are valid at all times during the dynamics, giving, for example, a qubit's fidelity and entropy as its wavepacket flies through the interaction region, as is done in Fig.~\ref{fig:FidelityEntropyPlot}, which shows that $F (t)$ and $S (t)$ are both non-monotonic functions of time.
Eq.~\eqref{eq:InternalState} implies that if $\rho_0 \propto \Id$ then $\rho_I(t)\propto \Id$ at any time $t$, i.e., the dynamics is unital. This implies that the reservoir-free decoherence is non-Markovian~\cite{Das2018Fundamental}. A short but rigorous proof of this statement is given in sec. \ref{APPSec:Entropy} of the SM.

Notice that, in general, $F (t) < 1$ and $S (t) > 0$ while the particle is inside the interaction region even if the gate is perfectly implemented, such as the PHASE and cPHASE gates discussed below.

\vskip 1mm
\emph{Example with ballistic qubits.---}Let $H_0 = \frac{1}{2}\hbar \omega_q\sigma_z$, where $\sigma_z$ is the usual Pauli operator, and $\omega_q$ is the qubit frequency. We consider that the wavepacket entirely passes through an interaction region whose ideal dynamics perform a desired gate operation at final time $t_f$, where we recall that $H_{\rm NA}(t_f) = H_0$ and the typical energy scale is $E_0 = \hbar \omega_q$. We then evaluate the effect of the reservoir-free decoherence caused by the wavepacket's spatial spread. Notice that there is an infinite number of possible potentials $V(\hat{x})$ which would ideally implement a specific gate at final time. However, neither the final correction term, nor the final fidelity and entropy depend on this specific choice.

As first example, let us consider a NOT-gate, with ideal dynamics given by $U_{\rm NA} (t_f) = -i\sigma_x$, acting on an initial state $\rho_0= |\psi_I(0)\rangle\langle \psi_I(0)|$, with $|\psi_I(0)\rangle = \sqrt{a_0} |0\rangle + e^{i \theta}\sqrt{a_1} |1\rangle$, in the eigenbasis of $H_0$. Then, $\mcC(t_f) = - 2 \prtq{e^{i \theta}\sqrt{a_0 a_1}\dyad{0}{1} + \hc}$, 
and
\begin{equation}
\label{eq:fidelity-NOT-gate-example}
F (t_f) = 1 - K(t_f),
\quad
S (t_f) = K (t_f) (1 -\ln[K (t_f)])
\end{equation}
where $K (t_f) \equiv 4 a_0 a_1 (\omega_q \Delta x/v_0)^2$. As expected, the wavepacket's spatial spread reduces the gate fidelity, while increasing the qubit's entropy, for any initial state except eigenstates of $H_0$.

As second example, let us consider a PHASE-gate, whose ideal dynamics are $U_{\rm NA} (t_f) = \exp\big[-i(\phi/2)\sigma_z\big]$. Then $[U_{\rm NA} (t_f), H_0]=0$,
which implies $F (t_f)=1$ and $S (t_f)=0$ for all choices of $\rho_0$. 
Although entanglement is built during the interaction, the internal state associated with each position in the wavepacket undergoes the same dynamics once the wavepacket has completely passed through the interaction region. In other words, $\tU_x (t_f) = U_{\rm NA} (t_f), \forall\ x$ [cf. Eq.~\eqref{eq:UnitaryEvolutionPosition}].

More generally, an arbitrary gate operation has fidelity and entropy of the form given in Eq.~(\ref{eq:fidelity-NOT-gate-example}), but the quantity $K$ will be given by $(\omega_q \Delta x/v_0)^2$ multiplied by a prefactor which will depend on the gate operation (given by $U_{\rm NA}$) as well as the initial state, $\rho_0$.

\vskip 1mm
\emph{Two-qubit gate example.---} Our Eqs.~\eqref{eq:StartingHamiltonian} to \eqref{eq:EntropyPerturbation} can also describe the dynamics of two flying systems when their interaction only depends on their distance and we neglect the center of mass dynamics, see also sec.~V of the SM~\cite{}. Therefore, we can consider two flying qubits (1 and 2) traveling at different velocities along parallel 1D tracks. 
As one qubit flies past the other, their interaction $V(\abs{\hx_1 - \hx_2})$ performs a desired gate operation between them.
A cPHASE gate is unaffected by the reservoir-free decoherence, because the proper evolution of each qubit (under $H_1$ and $H_2$) commutes with the gate operation, so $\mcC (t_f) =0$. However, the cNOT gate is affected by noise scaling as $p (\Delta x_1^2 + \Delta x_2^2)/[\hbar^2 (v_1 - v_2)^2]$ (see sec.~\ref{APPSec:TwoFlyingSyst} of the SM) where $p$ is the population of the control qubit and we assumed the two qubit spatial states to be initially uncorrelated. The $\mcC (t_f)$ term has the same form as for the NOT gate [see above Eq.~\eqref{eq:fidelity-NOT-gate-example}] when the control qubit is in state $\ket{1}$ and is zero otherwise. The fidelity is then given by $F=1-pK$ where $K$ is defined as before but now refers to the qubit on which the NOT part of the gate acts. The entropy  is easily computed, but its formula is more involved and not given here.

\vskip 1mm
\emph{Experimental consequences.---}Ballistic electrons can be injected into quantum hall edge states 
on demand (Levitons, etc), and made to interact \cite{Dubois2013Oct,Freulon2015Apr,Kataoka2016Mar,Ferraro2018Dec}. They typically have $\Delta x /v_0\sim 10^{-10}$\,s \cite{Freulon2015Apr,Kataoka2016Mar}. If the electron's spin were used as a qubit, one would have $\omega_q \sim 10^{-10}$\,s$^{-1}$, since the B-fields $\gtrsim$ 1\,T. Then, the reservoir-free decoherence would be strong, $\omega_q\Delta x/v_0\sim 1$, giving fidelities much too small for quantum gate operations.

Achieving higher fidelities would require lower magnetic fields to get smaller $\omega_q$. This might be experimentally realizable with
electrons flying ballistically in a waveguide similar to \cite{Yamamoto2012Electrical} with B-fields of mT, allowing $\omega_q\sim 10^{-13}$\,s$^{-1}$. If the injection into this waveguide could be done with a similar $\Delta x$ as the injection into an edgestate, extremely high fidelities could be attained, with $1-F\sim (\omega_q\Delta x/v_0)^2\sim 10^{-6}$.

\vskip 1mm
\emph{Avoiding reservoir-free decoherence.---}The reservoir-free decoherence depends on the fact that, for each spatial point in the traveling wavefunction, the internal state experiences a different dynamics. However, if the initial internal state is an eigenstate of the bare Hamiltonian, $H_0$ (such as the ground state) it does not evolve prior to entering the interaction region and, once out of it, its evolution does not change the populations of states in the eigenbasis of $H_0$. Thus, the final internal state may be decohered in this energy eigenbasis, but the eigenstate's populations are the same as for the ideal dynamics.

This implies that reservoir-free decoherence plays no role whenever the system starts in an eigenstate of $H_0$, and it flies through potentials that (i) rotate the internal state to the desired superposition, (ii) perform a series of gate operations, and (iii) prepare the final state for an energy eigenbasis measurement~\cite{note_internal_state_obs}.
This is the case
in experiments on flying Rydberg atoms \cite{Najera-Santos2020Jul}, or flying electrons in waveguides \cite{Yamamoto2012Electrical}, explaining their negligible decoherence despite their wavepackets' large spatial spreads.

\begin{figure}
\centering
\includegraphics[width=0.48\textwidth]{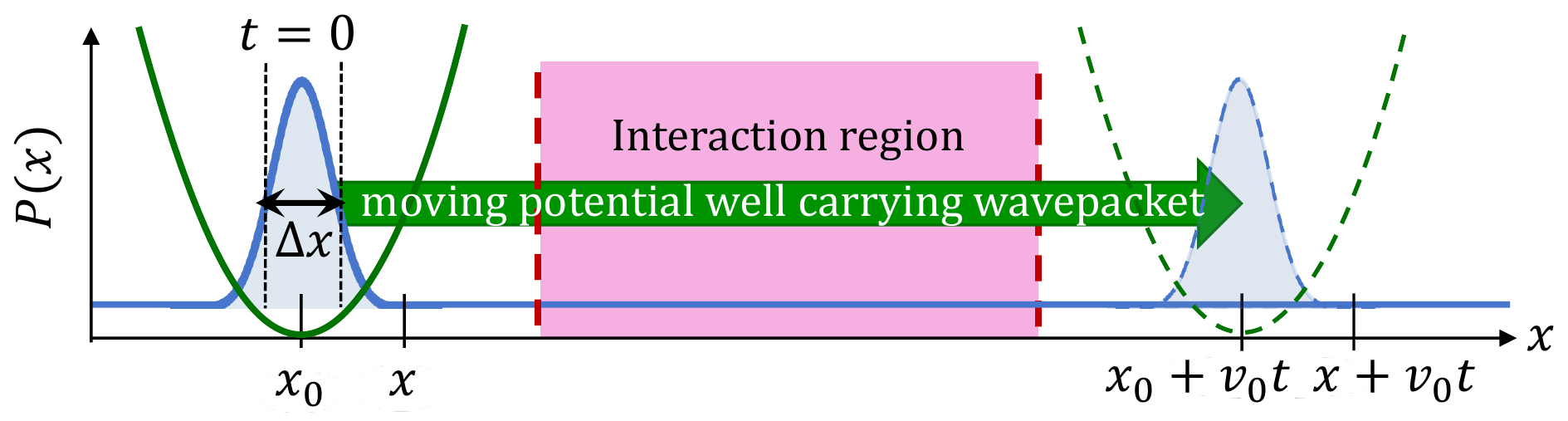}
\caption{A flying quantum system that is trapped in a potential well that moves at constant velocity, carrying the quantum system through the interaction region.
}
\label{fig:Illustration2}
\end{figure}

\vskip 1mm
\emph{Qubits carried by a moving potential.---}Finally, we consider qubits that fly by being trapped in a moving harmonic potential well~\cite{note_regime_harmonic}
, as in Fig.~(\ref{fig:Illustration2}); sometimes called flying qubits and sometimes called surfing or shuttling qubits \cite{Jadot2021Distant,Seidler2022Aug,Noiri2022Sep,Edlbauer2022semiconductor}. 
Sec.~\ref{APPSec:TrappingPotential} of the SM shows that the quantization of the wavefunction in the moving harmonic potential vastly reduces the reservoir-free decoherence compared to the ballistic qubits that are the principle subject of this letter.
Intuitively, this can be understood semi-classically: the particle undergoes a harmonic motion around the center of the moving trap, effectively averaging out the differences in internal dynamics in different parts of the wavepacket. Therefore all parts of the wavepacket have internal dynamics closer to the wavepacket's center than in the ballistic case.
As a result, the order $\delta x^2$-term in  Eq.~\eqref{eq:ApproximateDensityMatrixIDoFExplicit} is replaced by a term which we estimate to be smaller than $36(m^2/\hbar^2 v_0^2 \tau^4)\Delta x^6$, where $\tau$ is the time needed to apply the desired gate \cite{scaling}
. This term is of order $10^{-8}$ or smaller in most experiments (see sec.~\ref{APPSec:TrappingPotential_Magnitude} of the SM), which is small enough to completely neglect. 
Hence, reservoir-free decoherence can be effectively removed by switching from ballistic qubits to qubits carried by a moving harmonic potential wells.

\vskip 1mm
\emph{Conclusions.---}We considered flying quantum systems as bipartite systems divided into spatial and internal DoF, focusing on the case of ballistic particles with a narrow spatial distribution. We analytically investigated the effects of the wavepacket's spatial spread on the internal state dynamics, which experiences reservoir-free decoherence (decoherence without an external reservoir) due to the entanglement between spatial and internal DoF. We derived the internal state full dynamics, which is necessary in quantum thermodynamics, if one wants to go beyond our calculations of entropy, and quantify generalized work and heat using definitions like in Refs.~\cite{Weimer2008Local,HosseinNejad2015Work}. Finally, we estimated this effect to be practically negligible for surfing or shuttling qubits.
In the future, it would be interesting to explore how tuning the shape of the potential can reduce even more the reservoir-free decoherence, which fundamentally affects every flying quantum system.

\vskip 1mm
\emph{Acknowledgements.---}This work was supported by the John Templeton Foundation (Grant No. 61835). N. P. acknowledges the support by PNRR NQSTI Spoke 1 “Foundations and architectures for quantum information processing and communication” CUP: J93C22001510006. L. B. acknowledges the support by PNRR MUR project PE0000023-NQSTI. R.W. acknowledges the ANR Project “TQT’’ (ANR-20-CE30-0028). A. A. acknowledges the National Research Foundation, Singapore and A*STAR under its CQT Bridging Grant, the Plan France 2030 through the projects NISQ2LSQ ANR-22-PETQ-0006 and OQuLus ANR-23-PETQ-0013, ANR Research Collaborative Project “Qu-DICE” (Grant No. ANR-PRC-CES47) and the Foundational Questions Institute Fund (Grants No. FQXi-IAF19-01 and No. FQXi-IAF19-05). A. A. and R.W. acknowledge the ANR Research Collaborative Project “QuRes" (Grant No. ANR-PRC-CES47-0019).


\clearpage
\onecolumngrid
\begin{center}
	\textbf{\large Supplemental Material\\		
		Reservoir-free decoherence in flying qubits}
\end{center}

\setcounter{equation}{0}
\setcounter{figure}{0}
\setcounter{table}{0}
\setcounter{page}{1}
\setcounter{secnumdepth}{1}
\makeatletter
\renewcommand{\theequation}{S\arabic{equation}}
\renewcommand{\thefigure}{S\arabic{figure}}

\section{General analytical solution of the Clock Hamiltonian\label{APPSec:ClockHamiltonianSolution}}

Here, we analytically solve the \Schr equation based on the Clock Hamiltonian of the main text, which is
\begin{equation}
\label{APPeq:MasslessHamiltonian}
H = v_0 \hq + H_0 + V(\hx),
\end{equation}
where $\hq$ and $\hx$ are, respectively, the momentum and position operators for the spatial degree of freedom of the flying particle, $v_0$ is the constant speed at which the particle is moving, $H_0$ is the bare Hamiltonian of the internal degrees of freedom, and $V(\hx)$ is the interaction term which depends on the position operator $\hx$. The term $V(\hx)$ has the explicit form $V(\hx)=\sum_i f_i(\hx)V_i$, where the $f_i(x)$ are functions of $\hx$ and the $V_i$ are operators acting on the internal Degrees of Freedom (DoF) Hilbert space.

To solve the \Schr associated to the time-independent Hamiltonian of Eq.~\eqref{APPeq:MasslessHamiltonian}, we search for the propagator of the dynamics $\mcK (x,t;y) = \mel{x}{U (t)}{y}$, where $U (t)$ is the unitary operator which solves the \Schr equation:
\begin{equation}
i \hbar \dv{t} U(t) = H U(t),
\qq{with}
U(0) = \Id,
\end{equation}
where we also assumed $t_0 = 0$ and we omitted it. Then, we project on the left onto $\ket{y}$ and on the right onto $\bra{x}$, obtaining
\begin{equation}
i \hbar \dv{t} \mcK (x,t;y) = \prtq{-i\hbar v_0 \dv{x} + \tH(x)}\mcK (x,t;y),
\qq{where}
\tH (x) \equiv H_0 + V(x).
\end{equation}
given that $V(x)= \text{Tr}_{K}(V(\hx))=\sum_i f_i(x)V_i$, where the trace is over the spatial degree of freedom.
Notice that $\mcK (x,t;y)$ is still an operator in the Hilbert space of the internal DoF.

We now assume that the propagator is of the form
\begin{equation}
\label{APPeq:Propagator}
	\mcK (x,t;y) = \delta (x - (y + v_0 t)) \tU (x, y).
\end{equation}
Plugging this into the \Schr equation we are left with
\begin{equation}
i \hbar v_0  \delta (x - (y + v_0 t)) \dv{x} \tU (x, y) 
=  
\delta (x - (y + v_0 t)) \tH (x) \tU (x,y),
\end{equation}
which we can integrate over $x$ in order to obtain
\begin{equation}
	i \hbar \dv{t} \tU (y + v_0 t, y) =  \tH (y + v_0 t) \tU (y + v_0 t, y).
\end{equation}
where we recall that $\dd{x}= v_0 \dd{t}$ after the integration.
Hence, the solution to the above equation is
\begin{equation}
\label{APPeq:NonAutonomousUnitaryOperator}
	\tU (y + v_0 t, y) = \mathcal{T} \exp[- \frac{i}{\hbar} \int_{0}^{t} \dd{s} \tH (y + v_0 s)], 
\end{equation}
where $\mathcal{T}$ is the time ordering operator. The above formula is also the formal solution of the non-autonomous model for a point-like flying particle.
Therefore, we have proven that the propagator given in Eq.~\eqref{APPeq:Propagator} solves the time-dependent \Schr equation associated to the time-independent Hamiltonian of Eq.~\eqref{APPeq:MasslessHamiltonian}. 

The most generic state at time $t=0$ can be written as $\mel{x}{\rho(0)}{y} = A_0 (x,y) \rho_0 (x,y)$, where $\ket{x}$ and $\ket{y}$ are position eigenstates in the spatial degree of freedom (DoF) Hilbert space and $\rho_0 (x,y)$ is an operator in the Hilbert space of the internal DoF at spatial points $(x,y)$. We take $A_0 (x,x) \geq 0$ and $\rho_0 (x,x)$ to be a proper density matrix so that $A_0 (x,x)$ represents the probability density function of finding the particle at point $x$ at $t=0$. In the case when the initial state is a product state, $\rho_0 (x,y)$ does not depend on $(x,y)$ and one can simply write $\rho_0$. Moreover, if the initial state of the spatial degree of freedom is pure, $A_0 (x,y) = \psi_0 (x) \psi_0^* (y)$, where $\psi_0 (x)$ is the wavefunction of the spatial degree of freedom at time $t=0$.
Finally, at time $t$, the state can be written as:
\begin{equation}
\label{APPeq:GeneralDynamics}
\mel{x+v_0 t}{\rho (t)}{y + v_0 t} = A_0 (x,y) U_x (t) \rho_0 (x,y) U_y^\dg (t),
\end{equation}
where we defined $\tU_x (t) \equiv \tU (x + v_0 t, x)$.

Starting from Eq.~\eqref{APPeq:GeneralDynamics}, the reduced density operator for the spatial degree of freedom is
\begin{equation}
\label{APPeq:KDoFReducedDensityMatrix}
\mel{x+v_0 t}{\rho_K (t)}{y+ v_0 t}
=
A_0 (x,y) \Tr \prtq{U_x (t) \rho_0 (x,y) U_y^\dg (t)},
\end{equation}
while the reduced density matrix of the internal degree of freedom is
\begin{eqnarray}
\rho_I (t) \equiv \int_{-\infty}^{+\infty} \dd{x} \mel{x}{\rho (t)}{x}
=  \int_{-\infty}^{+\infty} \dd{x} A_0 (x,x) U_x (t) \rho_0 (x,x) U_x^\dg (t).
\label{APPeq:DensityMatrixIDoF}
\end{eqnarray}
Notice that $\mel{x}{\rho_K (t)}{x} = A_0 (x-v_0 t,x - v_0 t)$, i.e., the position probability density function travels at constant velocity $v_0$ independently of everything else. This is in accord with the fact that, in Heisenberg picture, one can immediately write $\hx_H (t) = \hx_H (0) + v_0 t$. Finally, let us also notice that the when $\rho_0 (x,x) \propto \Id$, then $\rho_I (t) \propto \Id$, implying that the map governing the dynamics of the internal degree of freedom  is unital at all times.

When the initial state is pure, $\rho (0) = \dyad{\psi (0)}$, with $\ket{\psi(0)}=\intmp \dd{x} \Psi_0 (x)\ket{\psi_I (0)}$, we get that $\rho_0(x,x)= \ket{\psi_I (0)}\bra{\psi_I (0)}$ and $A_0(x,x) = |\Psi_0 (x)|^2$. Then, the state at time $t$ and position $x+v_0 t$ is  $\Psi_0 (x) \tU_x (t)\ket{\psi_I (0)}$. This is the result reported in the main text, in the paragraph containing Eq.~(3).


\clearpage
\section{Perturbative expansion for localized wavepackets\label{APPSec:LocalizedWavePackets}}

In this section, we make use of localized wavepackets in order to derive simplified expressions for the dynamics of the system described by Eq.~\eqref{APPeq:GeneralDynamics}.
First, we notice that the non-autonomous dynamics of the internal degree of freedom associated to the unitary operator $\tU(x+v_0 t, x)$ can always be written as
\begin{equation}
\tU (x+v_0 t, x) = \tU (x+v_0 t, x_0 + v_0 t) \tU (x_0 + v_0 t , x_0) \tU^\dg (x, x_0).
\end{equation}
Let us start by considering the second order expansion of the Dyson's series of $\tU (x+v_0 t, x_0 + v_0 t)$ [see Eq~\eqref{APPeq:NonAutonomousUnitaryOperator}]:
\begin{equation}
\label{APPeq:DysonSeriesSecondOrder}
\tU (x+v_0 t, x_0 + v_0 t) \simeq \Id - \frac{i}{\hbar}\int_{0}^{t'} \dd{s} \tH \prt{x_0 + v_0 (t + s)} - \frac{1}{2 \hbar^2} \int_{0}^{t'} \int_{0}^{t'} \dd{s} \dd{s'} \tH \prt{x_0 + v_0 (t + s)} \tH \prt{x_0 + v_0 (t + s')},
\end{equation}
where $t'= \delta x/v_0$ and $\delta x = x-x_0$.
We consider the wavepacket to be localized also with respect to the spatial variations of $\tH \prt{x_0 + v_0 (t + s)}$ so that we can write $\tH \prt{x_0 + v_0 (t + s)} \simeq \tH \prt{x_0 + v_0 t} + v_0 s \partial_{x} \tH \prt{x}|_{x_0 + v_0 t}$. Substituting this back in Eq.~\eqref{APPeq:DysonSeriesSecondOrder} we get
\begin{equation}
\tU (x+v_0 t, x_0 + v_0 t) \simeq \Id - i \frac{\delta x}{\hbar v_0} \tH_f - \frac{1}{2} \prt{\frac{\delta x}{\hbar v_0}}^2 \tH_f^2 - \frac{i \hbar v_0}{2} \prt{\frac{\delta x}{\hbar v_0}}^2 \tH_f'.
\end{equation}
where we defined $\tH_f \equiv \tH (x_0 + v_0 t)$ and $\tH_f' \equiv \partial_{x} \tH (x) |_{x_0 + v_0 t}$. In an analogous manner, we can obtain the second order expansion for $\tU^\dg (x,x_0)$:
\begin{equation}
\tU^\dg (x,x_0) \simeq \Id + i \frac{\delta x}{\hbar v_0} \tH_p - \frac{1}{2} \prt{\frac{\delta x}{\hbar v_0}}^2 \tH_p^2 + \frac{i \hbar v_0}{2} \prt{\frac{\delta x}{\hbar v_0}}^2 \tH_p',
\end{equation}
where we defined $\tH_p \equiv \tH (x_0)$ and $\tH_p' \equiv \partial_{x} \tH (x) |_{x_0}$.
If the initial wavefunction is well localized around $x_0$, we can approximate $\tU (x+v_0 t, x)$ as follows:
\begin{equation}
\label{APPeq:ApproximatedUnitaryNA}
\tU (x+v_0 t, x) \simeq U_{\rm NA} (t) - i\frac{\delta x}{\hbar v_0} U_{1} (t) + \prt{\frac{\delta x}{\hbar v_0}}^2 U_{2} (t),
\end{equation}
where
\begin{equation}
U_{1} (t) = \tH_f U_{\rm NA} - U_{\rm NA} \tH_p,
\quad
U_{2} (t) = \tH_f U_{\rm NA} \tH_p
- \frac{1}{2}\prt{\tH_f^2 U_{\rm NA} + U_{\rm NA} \tH_p^2} - \frac{i \hbar v_0}{2}\prt{\tH_f' U_{\rm NA} - U_{\rm NA} \tH_p'}.
\end{equation}
where we suppressed time dependencies on the right-hand-side of the equations. Notice that the operators $U_{n} (t)$ defined above do not depend on the position $x$ but only on $x_0$. In a scattering dynamics, the Hamiltonians long before and after the interaction are the same and are constant. Therefore, we can write $\tH_f \simeq \tH_p \simeq H_0$ and $\tH_f' \simeq \tH_p' \simeq 0$. It follows that the operators can be simplified as follows
\begin{equation}
U_{1} (t) = \comm{H_0}{U_{\rm NA} (t)},
\quad
U_{2} (t) = H_0 U_{\rm NA} (t) H_0 - \frac{1}{2}\acomm{H_0^2}{U_{\rm NA} (t)}.
\label{APPeq:U1U2}
\end{equation}
Substituting Eq.~\eqref{APPeq:ApproximatedUnitaryNA} into Eq.~\eqref{APPeq:GeneralDynamics} and keeping only second order terms, the global state at time $t$ assumes the following form:
\begin{multline}
\label{APPeq:WavepacketDynamicsDensityOperatorSecondOrder}
\mel{x+v_0 t}{\rho (t)}{y+ v_0 t}
\simeq 
A_0 (x,y) \prtqB{U_{\rm NA} \rho_0 (x,y) U_{\rm NA}^\dg +\frac{i}{\hbar v_0}\prt{\delta y U_{\rm NA} \rho_0 (x,y) U_1^\dg - \delta x U_1 \rho_0 (x,y) U_{\rm NA}^\dg} +
\\
+\frac{1}{\hbar^2 v_0^2} \prt{\delta y^2 U_{\rm NA} \rho_0 (x,y) U_2^\dg + \delta y \delta x U_1 \rho_0 (x,y) U_1^\dg + \delta x^2 U_2 \rho_0 (x,y) U_{\rm NA}^\dg}},
\end{multline}
where we omitted time dependencies in the right hand side of the equation and introduced the symbol $\delta y \equiv y - x_0$.

\clearpage
\section{Reduced state of the internal degree of freedom\label{APPSec:Reducedstate}}

Here, we derive the approximate form that the density operator of the internal degree of freedom takes under the assumption that internal and spatial degrees of freedom are initially uncorrelated and that the spatial wavepacket is narrow. Under these assumptions, we trace Eq.~\eqref{APPeq:WavepacketDynamicsDensityOperatorSecondOrder} over the spatial degree of freedom obtaining
\begin{multline}
\label{APPeq:WithU1U2}
\rho_I (t) 
\simeq U_{\rm NA} (t) \rho_0 U_{\rm NA}^\dg (t) \prtq{\int_{-\infty}^{+\infty} \dd{x} A_0 (x,x)}
+\frac{i}{\hbar v_0}\prtq{U_{\rm NA} (t) \rho_0 U_1^\dg (t) - U_1 (t) \rho_0 U_{\rm NA}^\dg (t)} \prtq{\int_{-\infty}^{+\infty} \dd{x} A_0 (x,x) \delta x}
\\
+\frac{1}{\prt{\hbar v_0}^2}
\prtq{U_{\rm NA} (t) \rho_0 U_2^\dg (t) + U_1 (t) \rho_0 U_1^\dg (t) + U_2 (t) \rho_0 U_{\rm NA}^\dg (t)}
\prtq{\int_{-\infty}^{+\infty} \dd{x} A_0 (x,x) \prt{\delta x}^2}. 
\end{multline}
The integral for the zero order term is equal to one and the integral for the first order term is equal to zero since $x_0$ is the mean position at $t=0$ by definition. The integral for the second order term is instead, again by definition, $(\Delta x)^2$, i.e., the dispersion in position of the wavepacket. Therefore, the approximate density matrix for the internal degree of freedom at time $t$ is given by
\begin{equation}
\label{APPeq:ApproximateDensityMatrixIDoFUnitaries}
\rho_I (t) 
\simeq U_{\rm NA} (t) \rho_0 U_{\rm NA}^\dg (t) 
+\prt{\frac{\Delta x}{\hbar v_0}}^2
\prtq{U_1 (t) \rho_0 U_1^\dg (t) + U_{\rm NA} (t) \rho_0 U_2^\dg (t) + U_2 (t) \rho_0 U_{\rm NA}^\dg (t)}.
\end{equation}
The above equation can be rewritten explicitly in terms of the non-autonomous evolution given by $U_{\rm NA} (t)$ and the Hamiltonian $\tH(x_0 + v_0 t)$ as
\begin{multline}
\label{APPeq:ApproximateDensityMatrixIDoFExplicit}
\rho_I (t) 
\simeq U_{\rm NA} \rho_0 U_{\rm NA}^\dg 
+\prt{\frac{\Delta x}{\hbar v_0}}^2
\prtgB{
- \frac{i \hbar v_0}{2}\prt{\comm{\tH_f'}{U_{\rm NA} \rho_0 U_{\rm NA}^\dg} - U_{\rm NA}\comm{\tH_p'}{\rho_0}U_{\rm NA}^\dg} + \\
U_{\rm NA} D_{\tH_p} (\rho_0) U_{\rm NA}^\dg + D_{\tH_f} \prt{U_{\rm NA} \rho_0 U_{\rm NA}^\dg} + \comm{\tH_f}{U_{\rm NA} \comm{\tH_p}{\rho_0} U_{\rm NA}^\dg}},
\end{multline}
where we suppressed the time dependencies and used the general notation $D_{X} (\rho) = X \rho X^\dg - (1/2)\acomm{X^\dg X}{\rho}$. Finally, we can rewrite the Eq.~\eqref{APPeq:ApproximateDensityMatrixIDoFExplicit} in the following way:
\begin{equation}
\label{APPeq:ApproximateDensityMatrixIDoFCompact}
\rho_{\rm I} (t) \simeq \rho_{\rm NA} (t) + \prt{\frac{ \Delta x}{\hbar v_0}}^2 \mcC (\rho_0, t),
\end{equation}
where we defined the operator $\rC (t)$,  and defined the symbols $\rho_{\rm NA} (t) \equiv U_{\rm NA} \rho_0 U_{\rm NA}^\dg$, and
\begin{multline}
\mcC (\rho_0, t) \equiv  U_{\rm NA} D_{\tH_p} (\rho_0) U_{\rm NA}^\dg + D_{\tH_f} \prt{U_{\rm NA} \rho_0 U_{\rm NA}^\dg} 
+\\+ 
\comm{\tH_f}{U_{\rm NA} \comm{\tH_p}{\rho_0} U_{\rm NA}^\dg} - \frac{i \hbar v_0}{2}\prt{\comm{\tH_f'}{U_{\rm NA} \rho_0 U_{\rm NA}^\dg} - U_{\rm NA}\comm{\tH_p'}{\rho_0}U_{\rm NA}^\dg},
\end{multline}
which has the dimension of an energy squared. 
The subscript of $\rho_{\rm NA} (t)$ stands for \enquote{Non-Autonomous} while the symbol $\mcC$ stands for \enquote{Correction}.

The operator $\rC (t)$ has trace one and is Hermitian. One can easily check that $\mcC (\rho_0, t)$ is traceless from its definition, while it is easier to check that it is Hermitian from Eq.~\eqref{APPeq:ApproximateDensityMatrixIDoFUnitaries}. However, for $\rC (t)$ to be a proper density operator, it has also to be positive semi-definite, i.e., all its eigenvalues have to lie in the interval $[0,1]$. This is equivalent to say that $\ev{\rC (t)}{\psi} \geq 0$ for any ket $\ket{\psi}$. We expect that, as long as the perturbation is small enough, this condition is satisfied also in view of the fact that $\rC (t)$ comes from an approximation made on a proper density operator [see Eq.~\eqref{APPeq:DensityMatrixIDoF}].


\clearpage
\section{Entropy and Fidelity\label{APPSec:Entropy}}

The dynamics of the internal degree of freedom with respect to the ideal non-autonomous one can be characterized by many figures of merit. Two common ones are the von Neumann entropy and the fidelity. In this section and the next, we report the perturbative formulas for both of them. 

\subsection{Entropy and non-Markovianity}
In this subsection we provide the proof that a Markovian and unital dynamics leads to non-decreasing von Neumann entropy. It follows that a unital dynamics in which entropy can also decrease is non-Markovian.

A unital map $\mcE$ is a map such that $\mcE(\Id) = \Id$, i.e., the maximally mixed state is mapped onto itself. This is the case for the dynamics of Eq.~\eqref{APPeq:DensityMatrixIDoF} at all times. The contraction property of the relative entropy for quantum channels implies that $S(\mcE(\rho) || \mcE (\sigma)) \leq S(\rho || \sigma)$ where $\rho$ and $\sigma$ are arbitrary density matrices and $S(\rho || \sigma)$ is the relative entropy between them~\cite{Book_Nielsen2010}. The relative entropy between any state and the maximally mixed state is equal to $S(\rho || \Id/N) = \ln{N} - S(\rho)$, where $N$ is the Hilbert space dimension of the system and we recall that the possible highest von Neumann entropy for a system of dimensions $N$ is $S(\Id/N) = \ln{N}$. Finally, we recall that for a Markovian dynamics $\mcE_{t,t_0}$, bringing the state $\rho_0$ from $t_0$ to a later time $t$, it holds that $\mcE_{t,t_0} \rho_0 = \mcE_{t,t'} \mcE_{t',t_0} \rho_0$ for any $t_0 \leq t' \leq t$ because of the semi-group property~\cite{Book_Breuer2002}.
Under all these premises, we can now show that entropy cannot decrease in time:
\begin{equation}
S (\mcE_{t',t_0} \rho_0)
=
\ln{N} - S(\mcE_{t',t_0} \rho_0 || (\Id/N))
\leq 
\ln{N} - S(\mcE_{t,t'} \mcE_{t',t_0} \rho_0 || \mcE_{t,t'}(\Id/N))
=
S (\mcE_{t,t_0} \rho_0),
\implies
S (\mcE_{t,t_0} \rho_0) \geq S (\mcE_{t',t_0} \rho_0),
\end{equation}
for any $t \geq t' \geq t_0$.
It follows that dynamics considered in the main text is non-Markovian because it is unital, but 
its entropy can both increase and decrease in time. 

\subsection{Entropy}
The entropy of $\rC$ for small $\Delta x$ is given by the first order perturbative expansion of the entropy in Ref.~\cite{Grace2021Perturbation}, which extended previous derivations~\cite{XiaoYu2010Perturbation} that were only valid when $\rho_{\rm NA}$ is a full-rank density operator.
Even if the results are readily available in Ref.~\cite{Grace2021Perturbation}, in the following we explain how to obtain them by focusing on the physical relevance of the various regimes in which certain perturbations are allowed or not.

The perturbative treatment assumes the perturbation is small compared to the 
non-zero eigenvalues of the unperturbed density matrix, $\rho_{\rm NA}$. Of course, the perturbation cannot be treated as small compared to the zero eigenvalues of $\rho_{\rm NA}$, so we must treat them with care inside the perturbation expansion. This is critical, for instance, in applying the perturbation theory to experiments with flying particles, in which an accurate model of the intrinsic decoherence is most important when it is the dominant source of imperfections. Thus we are most interested in cases where the initial state is almost pure, and the imperfections arise due to the intrinsic decoherence.  In this case, we approximate the initial internal state as a pure state, $\rho_0=\dyad{\psi_{\rm NA} (t=0)}$ which means that the state at time $t$ is  $\rho (t)= \dyad{\psi_{\rm NA} (t)} + \prtq{\Delta x/(\hbar v_0)}^2 \mcC (\rho_0, t)$. In this case, the ideal state $\rho_{\rm NA} (t) = \dyad{\psi_{\rm NA} (t)}$ has $n-1$ zero eigenvalue (for an $n$-level internal state), which need careful treatment within the perturbation expansion. Hereafter, we drop the explicit time-dependencies to lighten the notation.

The opposite limit is one in which the initial internal state of the flying qubit is already strongly imperfect, and the intrinsic decoherence will only slightly increases the imperfection. This is of less interest for quantum information applications, which is why we do not discuss it in the main text, however it may be of use in understanding early experiments in which imperfections are likely to be common. In this case, it is likely that all the eigenvalues of the initial state $\rho_0$ are non-zero ($\rho_0$ is said to be full-rank) making the perturbation theory a little easier~\cite{XiaoYu2010Perturbation}.

To perturb about a $\rho_{\rm NA}$ with $m$ zero eigenvalues and $n-m$ non-zero eigenvalues, we rotate to the eigenbasis $\rho_{\rm NA}$, and organize the eigenstates such that non-zero eigenvalues of $\rho_{\rm NA}$ are in the first $n-m$ diagonal elements of the diagonalized matrix. Then we have  $\rho_{\rm NA} = \sum_{i=1}^{n-m} p_i \dyad{\phi_i}$ where $p_i > 0$ for $1\leq i \leq n-m$, and $p_i=0$ for $i>n-m$ (the $\ket{\phi_i}$s are orthonormal). 
We then break the perturbation $\mcC$ into blocks; the block $\mcC_\parallel$ acts on the states associated with non-zero eigenvalues of $\rho_{\rm NA}$ (the support of $\rho_{\rm NA}$) and the block  $\mcC_\perp$ acts on the states associated with the zero eigenvalues of $\rho_{\rm NA}$ (the kernel of $\rho_{\rm NA}$).
In other words, we define two projectors: $\Pi_{\parallel} = \sum_{i=1}^{n-m} \dyad{\phi_i}$ which projects onto the support of $\rho_{\rm NA}$, and $\Pi_{\parallel} = \sum_{i=n-m+1}^{n} \dyad{\phi_i}$ which projects onto the kernel of $\rho_{\rm NA}$. 
We then define $\mcC_{\parallel} \equiv \Pi_\parallel \mcC \Pi_\parallel$ and $\mcC_{\perp} \equiv \Pi_\perp \mcC \Pi_\perp$.
Then
\begin{equation}
\label{APPeq:EntropyPerturbationGeneral}
S(\rC) \simeq S(\rho_{\rm NA})  - \frac{\Delta x^2 }{\hbar^2 v_0^2}
\left(\Tr \Big(\mcC_\parallel + \mcC_\parallel \ln\big[\rho_{\rm NA}\big]\Big)  + \Tr\left(\mcC_\perp\ln
\left[\frac{\Delta x^2 \mcC_\perp}{\hbar^2 v_0^2}\right]\right) \right).
\end{equation}

If the initial state of the internal degree of freedom is a pure state, i.e., $\rho_0 = \dyad{\psi_0}$ it follows that the unperturbed state at time $t$ is $\rho_{\rm NA} = \dyad{\psi_{\rm NA}}$ where $\ket{\psi_{\rm NA}} = U_{\rm NA} \ket{\psi_0}$. Hence $S(\rho_{\rm NA})=0$ and $\mcC_\parallel= \ev{\mcC}{\psi_{\rm NA}}$, which means that
\begin{equation}\label{APPeq:EntropyPerturbationPureState}
S(\rC) \simeq -\frac{\Delta x^2 }{\hbar^2 v_0^2}
\left(\ev{\mcC}{\psi_{\rm NA}}  + \Tr\left(\mcC_\perp\ln
\left[\frac{\Delta x^2 \mcC_\perp}{\hbar^2 v_0^2}\right]\right) \right).
\end{equation}
This can be written in the compact form in the main text by
noting that $\Tr[\mcC]=0$ means $\ev{\mcC}{\psi_{\rm NA}}= - \Tr[\mcC_\perp]$.

%

In the opposite limit, when $\rho_{\rm NA}$ is full-rank, the result is simpler because $\mcC_{\parallel} = \mcC $ and $\mcC_{\perp} = 0$. Recalling that $\Tr [\mcC] = 0$, one immediately gets
\begin{equation}
\label{APPeq:EntropyPerturbationFullRank}
S(\rC) \simeq S(\rho_{\rm NA}) -  \frac{\Delta x^2 }{\hbar^2 v_0^2} \Tr\big(\mcC \ln(\rho_{\rm NA})\big).
\end{equation}

In general, the perturbation expansion used here works whenever the eigenvalues of the unperturbed density matrix divide into two categories; very small and large (the large ones being of order one). We can then choose the small parameter of the perturbation  expansion to be between the two, such that it is much less than the large eigenvalues (ensuring a controlled expansion), and much larger than the very small eigenvalues (which can be treated as being indistinguishable from zero).
It is this that allows a unique identification of 
the support and kernel of the matrix $\rho_{\rm NA}$. For example, let us consider the qubit density matrix $\rho_q = (1-q)\dyad{g} + q\dyad{e}$ and the perturbation $\delta \rho = \epsilon^2 \sigma_z$ with $\epsilon > 0$. The entropy of such a system is then $S (\rho_q + \epsilon^2 \sigma_z) = - (q+\epsilon^2)\ln(q+\epsilon^2) - (1-q-\epsilon^2)\ln(1-q-\epsilon^2)$. If we assume that $\epsilon^2 \ll q$ we get that $S (\rho_q + \epsilon^2 \sigma_z) \simeq S(\rho_q) - \epsilon^2 \ln(1/(1-q))$, in accordance with Eq.~\eqref{APPeq:EntropyPerturbationFullRank}. If, instead, we assume that $q \ll \epsilon^2$, we get that $S (\rho_q + \epsilon^2 \sigma_z) \simeq \epsilon^2 [1 - 2\ln(\epsilon)] - 2 q \ln(\epsilon)$, which agrees with Eq.~\eqref{APPeq:EntropyPerturbationPureState} upon setting $q=0$, i.e., treating $\rho_q$ as the pure state $\dyad{g}$.
When $q \sim \epsilon^2$, the approximated formula of Eq.~\eqref{APPeq:EntropyPerturbationGeneral} becomes unreliable; 
such cases must be treated by expanding in both $\epsilon$ and $q$, so Eq.~\eqref{APPeq:EntropyPerturbationGeneral} is the zeroth order term in an expansion in small $q=0$, and finite $q$ corrections must be calculated. However, we do not do that here.


\subsection{Fidelity\label{APPSec:Fidelity}}
To  the first order in $\Delta x^2$, the fidelity between $\rho_{\rm NA}$ and $\rC$ given in Eq.~\eqref{APPeq:ApproximateDensityMatrixIDoFCompact} is 
\begin{equation}
\label{APPeq:FidelityApproximatedFormula}
F = \prtq{\Tr \left(\sqrt{\sqrt{\rho_{\rm NA}}\ \rC \ \sqrt{\rho_{\rm NA}}}\right)}^2 \simeq 1 + \frac{\Delta x^2}{\hbar^2 v_0^2}\Tr \prtq{\mcC_\parallel}.
\end{equation}
When the initial state is pure, this reduces to: 
\begin{equation}
\label{Quad_F}
F = \ev{\rC}{\psi_{\rm NA}} \simeq 1 + \frac{\Delta x^2}{\hbar^2 v_0^2} \ev{\mcC}{\psi_{\rm NA}}.
\end{equation}

Interestingly, in the case when $\rho_{\rm NA}$ can be considered full-rank, the correction of order $\delta x^2$ to the fidelity vanishes in Eq.~\eqref{APPeq:FidelityApproximatedFormula}. This happens because $\mcC_{\parallel}=\mcC$, and so $\Tr[\mcC_{\parallel}]=\Tr[\mcC]=0$. Then, the deviation of the fidelity from one is of order $\delta x^4$. This means that if the initial state is already highly imperfect (i.e. far from pure), the reduction of fidelity induced by a small amount of intrinsic decoherence is much smaller than if the initial state is almost perfect (i.e.~initially close enough to pure that the intrinsic decoherence is the dominate source of impurity at time $t$). 

For completeness, we note that if $\rho_{\rm NA}$ is full-ranked, and $\rC=\rho_{\rm NA}+ (\Delta x^2\mcC)/(\hbar^2v_0^2)$, then we can work in the eigenbasis of $\rho_{\rm NA}$ to show that
the $i$th eigenvalue of 
$\sqrt{\rho_{\rm NA}}\,\rC\,\sqrt{\rho_{\rm NA}}$ is
\begin{eqnarray}
p_i^2 + p_i \frac{\Delta x^2}{\hbar^2v_0^2}\mcC_{ii} + 
\frac{\Delta x^4}{\hbar^4v_0^4}\sum_{j\neq i}\frac{p_ip_j |\mcC_{ij}|^2}{p_i^2-p_j^2}  + {\cal O}\left[\left(\frac{\Delta x^2 \mcC}{\hbar^2v_0^2}\right)^{\!\!3}\ \right]
\end{eqnarray}
where $p_i$ is the $i$th eigenvalue of $\rho_{\rm NA}$, and $\mcC_{ij}$ is the $ij$th element of $\mcC$ when written in the eigenbasis of $\rho_{\rm NA}$. 
Then the sum of the square roots of these eigenvalues gives $\Tr \left(\sqrt{\sqrt{\rho_{\rm NA}}\ \rC \ \sqrt{\rho_{\rm NA}}}\right)$, squaring this 
while 
using the fact that $\sum_i \mcC_{ii}=\Tr[\mcC]=0$, gives 
\begin{eqnarray}
F= 1 - \frac{\Delta x^4}{\hbar^4v_0^4}\sum_i \left(\frac{C_{ii}^2}{4p_i} -\sum_{j\neq i} \frac{p_j|C_{ij}|^2}{p_i^2-p_j^2} \right)
\end{eqnarray}
up to second order in the perturbation, i.e. up to lowest non-zero order in $\Delta x$.

\clearpage

\section{Two flying systems\label{APPSec:TwoFlyingSyst}}

We discuss here how Eq.~(1) of the main text applies to the case of two flying quantum systems with internal degrees of freedom, under the assumption that their interaction only depends on the distance between them. The Hamiltonian governing their dynamics is
\begin{equation}
    H = \frac{\hp_1^2}{2 m_1} + \frac{\hp_2^2}{2 m_2} + H_1 + H_2 + V (\hx_1 - \hx_2).
\end{equation}
The problem can be greatly simplified by moving to the reference frame of the center of mass~\cite{Book_Cohen2019QuantumMechanicsVol1}. The above Hamiltonian can be written as
\begin{equation}
H =  \frac{\hat{P}^2}{2 M} + \frac{\hp^2}{2 \mu} + H_0 + V(\hx).
\end{equation}
where we defined 
\begin{equation}
H_0 \equiv H_1 + H_2,
\quad
M \equiv m_1 + m_2,
\quad
\tm_i = \frac{m_i}{M}
\quad 
\mu \equiv \frac{m_1 m_2}{M},
\quad
\hat{P} \equiv \hp_1 + \hp_2,
\quad
\hp \equiv \tm_2 \hp_1 - \tm_1 \hp_2,
\quad
\hx \equiv \hx_1 - \hx_2.
\end{equation}
Since $\comm{\hat{P}}{\hp}=\comm{\hat{P}}{\hx}=0$ and $\comm{\hx}{\hp} = i \hbar$, we can neglect the dynamics of the mass center (whose momentum operator is $\hat{P}$) and focus on the remaining part, which has exactly the same form of the one-particle Hamiltonian. This particle has mass $\mu$, velocity $v_\mu \equiv \ev{p}/\mu = v_1 -v_2$, and mean position $\ev{\hx} = \ev{\hx_1} - \ev{\hx_2}$. Moreover, its spatial spread is given by
\begin{equation}
\Delta x^2 
= \ev{\hx^2} - \ev{\hx}^2
= \Delta x_1^2 + \Delta x_2^2 + 2 \prt{\ev{\hx_1}\ev{\hx_2}-\ev{\hx_1 \hx_2}},
\end{equation}
so that, if the two systems spatial wavefunctions are initially uncorrelated one gets $\Delta x = \sqrt{ \Delta x_1^2 + \Delta x_2^2}$. For completeness, we recall that the center of mass position operator is defined as follows $\hat{X} \equiv \tm_1 \hx_1 + \tm_2 \hx_2$~\cite{Book_Cohen2019QuantumMechanicsVol1}. 

If the narrow wavepacket approximation is not satisfied for the reduced mass wavefunction one needs to have its exact shape $A_r (x,x)$ in order to compute Eq.~\eqref{APPeq:DensityMatrixIDoF}. Notice that the reduced wavefunction is not needed because we still assume that we will be able to use the clock approximation to compute the dynamics of the internal states. We consider the initial state
\begin{equation}
    \rho (t=0) = \rho_{1,2} \otimes \prtq{\intmp \dd{x_1} \dd{x_2} \dd{y_1} \dd{y_2} A_{1,2} (x_1, x_2, y_1, y_2) \dyad{x_1,x_2}{y_1,y_2}},
\end{equation}
where $\rho_{1,2}$ is the density matrix of the internal state at $t=0$ and $A_{1,2} (x_1, x_2, y_1, y_2)$ is the spatial wavefunction. Since 
\begin{equation}
\ip{X,x}{x_1,x_2} = \delta \prt{x_1 - \prt{X + \tm_2x}} \delta \prt{x_2 - \prt{X - \tm_1x}},
\end{equation}
it follows that
\begin{equation}
    A_r (x,x) = \intmp \dd{X} A_{1,2} (X + \tm_2 x, X - \tm_1 x, X + \tm_2 x, X - \tm_1 x).
\end{equation}
When the two particles are initially uncorrelated and both in a pure state, the above equation becomes
\begin{equation}
    A_r (x,x) = \intmp \dd{X} \abs{\psi_1 (X + \tm_2 x)}^2 \abs{\psi_1 (X - \tm_1 x)}^2,
\end{equation}
where $\psi_1 (x)$ and $\psi_2 (x)$ are, respectively, the wavefunctions of particles one and two.

\textbf{Gaussian wavepackets:} In the case in which both particles have a gaussian profile and are uncorrelated at $t=0$, we get that the reduced mass spatial probability distribution is also a Gaussian. As computed before for the general case, this Gaussian has center $\ev{\hx} =\ev{\hx_1} - \ev{\hx_2}$ and spatial spread $\Delta x = \sqrt{ \Delta x_1^2 + \Delta x_2^2}$.

\clearpage
\section{Surfing or Shuttling particles\label{APPSec:TrappingPotential}}

In this section, we analyze the case in which a particle is moved around by moving the potential in which it is trapped (Notice that this moving potential is not the one acting on the internal DoF). We first derive a formula for the dynamics of the internal DoF due, again, to the spatial extension of the spatial DoF and then estimate the dependence of the correction to internal DoF dynamics to wavepacket spread in position for typical situations.

\subsection{Perturbation due to wavepacket spreading}

We study a particle that is trapped in a potential moving at speed $v_0$. If this is done perfectly, we can stay in the reference frame such that the center of the potential is always at $x=0$. The Hamiltonian of the trapped particle is therefore $H_K \equiv \hp^2/(2m) + W (\hx)$, where $W (x)$ represents the potential. For example, for a harmonic oscillator, this would be $W (x) = (1/2) m \omega^2 x^2$. In other words, $H_K$ is the Hamiltonian of the spatial DoF.

Now we assume that the particle also has internal DoF, with bare Hamiltonian $H_0$. While moving by means of the traveling potential, there is another potential $V(\hx)$ acting solely on the internal DoF. In the reference frame at rest with the trap, we get that
\begin{equation}
H 
= H_K + H_0 + V(v_0 t + \hx) \simeq H_K + H_0 + V(v_0 t) + \frac{\hx}{v_0}\dot{V} (v_0 t),
\end{equation}
where we assume that the spatial variation of this potential is much smaller than the width of the particle's trap so that the last term is a perturbation term. Since we treat $v_0$ as a parameter, henceforth we will write $V(t)$ in place of $V(v_0 t)$.

If there was no perturbation term, the  non-autonomous evolution would be as follows:
\begin{equation}
\mcU_{\rm NA} (t) = e^{-\frac{i t}{\hbar}H_K} U_{\rm NA} (t),
\qq{where}
U_{\rm NA} (t) = \mcT \prtg{ - \frac{i}{\hbar} \int_0^t \prtq{H_0 + V(s)} \dd{s}},
\end{equation}
where we exploited the fact that $H_K$ and $H_0 + V(t)$ act on different Hilbert spaces. One can now write the \Schr equation in interaction picture with respect to $H_K + H_0 + V(t)$. The interaction picture state is given by $\ket{\psi_I (t)} = \mcU_{\rm NA}^\dg (t) \ket{\psi (t)}$. The \Schr equation in interaction picture is
\begin{equation}
i \hbar \dv{t} \ket{\psi_I (t)} = H_I (t) \ket{\psi_I (t)},
\qq{where}
H_I (t) = \frac{1}{v_0} \mcU_{\rm NA}^\dg (t) \hx \dot{V} (t) \mcU_{\rm NA} (t).
\end{equation}
We denote the unitary solving the above \Schr equation by $\mcK (t)$ and using the Dyson series approach up to second order we have that
\begin{equation}
\mcK (t) =1 - \frac{i}{\hbar}\int_0^t H_I (s) \dd{s} - \frac{1}{\hbar^2}\int_0^{t} \dd{s_1}\int_0^{s_1} \dd{s_2} H_I (s_1) H_I (s_2). 
\end{equation}
It follows that the state at time $t$ in \Schr picture is approximately given by $\ket{\psi (t)} \simeq \mcU_{NA} (t) \mcK (t) \ket{\psi (0)}$.

Now we assume that the initial state is $\ket{\psi (0)} = \ket{g_K}\otimes \ket{\psi_0}$ where $\ket{g_K}$ is the ground state of $H_K$, and we choose its eigenvalue to be zero. We also assume that the only relevant transition happening on the spatial DoF is from $\ket{g_K}$ to the first excited level $\ket{e_K}$ with energy $\mcE_{eg}$. This means that for our evolution the only important \enquote{matrix elements} of $\mcK (t)$ are $\mel{g_K}{\mcK (t)}{g_K}$ and $\mel{e_K}{\mcK (t)}{g_K}$. In this approximation, we have that $\dyad{g_K} + \dyad{e_K} = \Id$. Moreover, we assume that $\ev{\hx}{g_K} = \ev{\hx}{e_K} = 0$, i.e., ground and excited states of the spatial DoF are centered.

For the first matrix element, we get
\begin{equation}
\mel{g_K}{\mcK (t)}{g_K} = 1 - \frac{1}{\hbar^2} \int_0^{t} \dd{s_1}\int_0^{s_1} \dd{s_2} \mel{g_K}{H_I (s_1) H_I (s_2)}{g_K},
\end{equation}
where we used the fact that $\ev{H_I (t)}{g_K} = 0$ at all times.
Then, we have
\begin{equation}
\mel{g_K}{H_I (s_1) H_I (s_2)}{g_K} =
\frac{1}{v_0^2}U_{\rm NA}^\dg (s_1) \dot{V}(s_1)
\ev{\hx \mcU_{\rm NA} (s_1) \mcU^\dg_{\rm NA} (s_2) \hx}{g_K}\dot{V}(s_2) U_{\rm NA} (s_2),
\end{equation}
where we used $\mcU_{\rm NA} (t) \ket{g_K} = \ket{g_K}U_{\rm NA} (t)$. Now, we can insert $\dyad{g_K} + \dyad{e_K}$ between the $\mcU_{\rm NA}$s and we get
\begin{equation}
\ev{\hx \mcU_{\rm NA} (s_1) \mcU^\dg_{\rm NA} (s_2) \hx}{g_K}
=\\
e^{- i \mcE_{eg} (s_1 - s_2)/\hbar} \abs{\mel{g_K}{\hx}{e_K}}^2 U_{\rm NA} (s_1) U_{\rm NA}^\dg (s_2).
\end{equation}
For the other matrix element, instead, we have
\begin{equation}
\mel{e_K}{\mcK (t)}{g_K} = 1 - \frac{i}{\hbar}\int_0^t \mel{e_K}{H_I (s)}{g_K} \dd{s},
\qq{where}
\mel{e_K}{H_I (s)}{g_K} = \frac{e^{ i \frac{s \mcE_{eg}}{\hbar}}}{v_0} \mel{e_K}{\hx}{g_K} U^\dg_{\rm NA} (s) \dot{V} (s) U_{\rm NA} (s).
\end{equation}

Putting everything together, when $\mcK (t)$ is applied to the state $\ket{g_K}\ket{\psi_0}$, we have that
\begin{equations}
\mcK (t) &\simeq 1 - i\dyad{e_K}{g_K}\frac{r_{eg}}{\hbar v_0}X_t^{(1)} - \frac{1}{2}\dyad{g_K}{g_K} \frac{\abs{r_{eg}}^2}{\hbar^2 v_0^2} X_t^{(2)},\qquad
X_t^{(1)} \equiv \int_0^t e^{ i \frac{s \mcE_{eg}}{\hbar}}U^\dg_{\rm NA} (s) \dot{V} (s) U_{\rm NA} (s) \dd{s},\\
X_t^{(2)} &\equiv 2\int_0^t \dd{s_1} \int_0^{s_1} \dd{s_2} e^{- i \mcE_{eg} (s_1 - s_2)/\hbar} U_{\rm NA}^\dg (s_1)\dot{V} (s_1) U_{\rm NA} (s_1) U_{\rm NA}^\dg (s_2) \dot{V} (s_2) U_{\rm NA} (s_2),
\end{equations}
where we defined $r_{eg} \equiv \mel{e_K}{\hx}{g_K}$. 

In interaction picture, the internal DoF state is given by
$\rho_I (t) \simeq \Tr_K \prtg{\mcK (t) \prtq{\dyad{g_K}\otimes \rho_0} \mcK^\dg (t)}$. By substituting what we previously found and keeping terms only up to second order in $r_{eg}/(\hbar v_0)$ we get
\begin{equation}
\label{APPeq:ShuttlingQubitIDoF}
\rho_I (t) \simeq \rho_0 - \frac{\abs{r_{eg}}^2}{2\hbar^2 v_0^2}\prtg{X_t^{(2)}\rho_0 + \rho_0 X_t^{(2),\dg} - 2 X_t^{(1)}\rho_0 X_t^{(1),\dg}}.
\end{equation}

\subsection{Estimation of ideal dynamics deviation}

In order to evaluate how high is the deviation from the non-autonomous dynamics, we want to estimate the magnitude of the operator $X_t^{(1)}$. We do this by considering the fact that the potential and the non-autonomous evolution happen on scales which are much longer than $\hbar/\mcE_{eg}$. We define $A(t) \equiv U^\dg_{\rm NA} (t) \dot{V} (t) U_{\rm NA} (t)$ and $t_n \equiv n \delta t$ with $\delta t =2 \pi \hbar/\mcE_{eg}$. We assume that we can expand to first order $A(t_n + t)$ for $\abs{t}\leq \delta t$ and we write
\begin{multline}
\norm{X_t^{(1)}}
\simeq
\norm{\int_0^{N \delta t} e^{ i \frac{t \mcE_{eg}}{\hbar}} A(t) \dd{t}}
\simeq
\norm{\sum_{n=0}^{N-1} \int_{t_n}^{t_{n+1}} e^{ i \frac{t \mcE_{eg}}{\hbar}} \prtq{A(t_n) + (t-t_n) \dot{A} (t_n)} \dd{t}}
=
\norm{-\frac{2 i \pi \hbar^2}{\mcE_{eg}^2} \sum_{n=0}^{N-1} \dot{A} (t_n)}\\
= \frac{2 \pi \hbar^2}{\mcE_{eg}^2}  \times
\norm{\sum_{n=0}^{N-1} U_{\rm NA}^\dg (t_n) \prtg{\frac{i}{\hbar}\comm{H(t_n)}{\dot{V} (t_n)} + \ddot{V} (t_n)} U_{\rm NA} (t_n)},
\end{multline}
where $N$ is the relevant number of steps in which the potential is not negligible. The norm considered is left unspecified for more generality and because its exact form is not relevant for the estimation. Due to the subadditivity property of norms~\cite{Book_Hall2013Quantum}, we can overestimate the above quantity by summing the norm of each element separately. Hence we can therefore eliminate the unitary operators sandwiching the Hermitian operators.
\begin{multline}
\label{APPeq:NormX1T}
\norm{X_t^{(1)}} 
\lesssim 
\frac{2 \pi \hbar^2}{\mcE_{eg}^2} \sum_{n=0}^{N-1} \norm{\frac{i}{\hbar}\comm{H(t_n)}{\dot{V} (t_n)} + \ddot{V} (t_n)}
\leq
\frac{2 \pi \hbar^2}{\mcE_{eg}^2} \sum_{n=0}^{N-1} 
\prtg{\frac{1}{\hbar}\norm{\comm{H(t_n)}{\dot{V} (t_n)}} + \norm{\ddot{V} (t_n)}}
\leq\\ 
\leq
\frac{2 \pi \hbar^2}{\mcE_{eg}^2} \sum_{n=0}^{N-1} 
\prtg{\frac{2}{\hbar}\norm{H(t_n)}\norm{\dot{V} (t_n)} + \norm{\ddot{V} (t_n)}},
\end{multline}
where we also exploited the property of norms $\norm{AB} \leq \norm{A}\norm{B}$~\cite{Book_Hall2013Quantum}. From there, we can assume that the gate lasts a time $\tau$, high enough to perform, for example, a qubit rotation. We then have $N \delta t \sim \tau$ and we write
\begin{equation}
\label{APPeq:EstimationGateTime}
\norm{H(t_n)} \sim \frac{\hbar}{\tau},\quad
\norm{\dot{V} (t_n)} \sim \frac{\hbar}{\tau^2},\quad 
\norm{\ddot{V} (t_n)} \sim \frac{\hbar}{\tau^3},
\implies
\norm{X_t^{(1)}} 
\lesssim 
\frac{2 \pi \hbar^2 N}{\mcE_{eg}^2} \frac{3 \hbar}{\tau^3}
=
\frac{3 \hbar^2}{\tau^2 \mcE_{eg}},
\end{equation}
where we exploited the relation $N = \tau/\delta t = \tau \mcE_{eg}/(2 \pi \hbar)$. Finally, we can estimate that, in the equation for $\rho_I (t)$, the second correction term has magnitude
\begin{equation}
\label{APPeq:MagnitudeX1}
\norm{\frac{\abs{r_{eg}}^2}{\hbar^2 v_0^2}X_t^{(1)}\rho_0 X_t^{(1),\dg}} 
\lesssim \frac{\abs{r_{eg}}^2}{\hbar^2 v_0^2}\frac{3 \hbar^2}{\tau^2 \mcE_{eg}}\frac{3 \hbar^2}{\tau^2 \mcE_{eg}}  
= \frac{9 \hbar^2}{v_0^2 \tau^4} \frac{\abs{r_{eg}}^2}{\mcE_{eg}^2}.
\end{equation}

The same kind of calculations can be performed to estimate the value of $\norm{X_t^{(2)}}$. We write
\begin{multline}
\norm{X_t^{(2)}}
\simeq
2 \norm{ \int_0^{N \delta t} \dd{s_1} \int_0^{s_1} \dd{s_2} e^{- i \mcE_{eg} (s_1 - s_2)/\hbar} A(s_1) A(s_2)}
\simeq\\
\simeq
2 \norm{\sum_{n=0}^{N-1} \sum_{m=0}^{n} 
\int_{t_n}^{t_{n+1}} \dd{s_1} \int_{t_m}^{t_{m+1}} \dd{s_2} e^{- i \mcE_{eg} (s_1 - s_2)/\hbar} \prtq{A(t_n) + (s1-t_n) \dot{A} (t_n)} \prtq{A(t_m) + (s2-t_m) \dot{A} (t_m)}}
=\\
=
\frac{8 \pi^2 \hbar^4}{\mcE_{eg}^4} \norm{\sum_{n=0}^{N-1} \sum_{m=0}^{n}  \dot{A}(t_n)\dot{A}(t_m)}
\leq
\frac{8 \pi^2 \hbar^4}{\mcE_{eg}^4} \sum_{n=0}^{N-1} \sum_{m=0}^{n}  \norm{\dot{A}(t_n)\dot{A}(t_m)}
\leq
\frac{8 \pi^2 \hbar^4}{\mcE_{eg}^4} \sum_{n=0}^{N-1} \sum_{m=0}^{n}  \norm{\dot{A}(t_n)}\norm{\dot{A}(t_m)}
\leq\\
\leq
\frac{8 \pi^2 \hbar^4}{\mcE_{eg}^4} \sum_{n=0}^{N-1} \sum_{m=0}^{n} \prtg{\frac{2}{\hbar}\norm{\prtg{H(t_n)}}\norm{\dot{V} (t_n)} + \norm{\ddot{V} (t_n)}}\prtg{\frac{2}{\hbar}\norm{H(t_m)}\norm{\dot{V} (t_m)} + \norm{\ddot{V} (t_m)}}
\sim\\
\sim
\prt{\frac{2 \pi \hbar^2 N}{\mcE_{eg}^2} \frac{3 \hbar}{\tau^3}}^2
=
\prt{\frac{3 \hbar^2}{\tau^2 \mcE_{eg}}}^2,
\end{multline}
where we notice that the estimate gives exactly the same result as in Eq.~\eqref{APPeq:EstimationGateTime}. It follows that the entire deviation from the ideal dynamics scales, at worst, as $9 \hbar^2 \abs{r_{eg}}^2/(v_0^2 \tau^4 \mcE_{eg}^2)$.
In other words, the worst-case infidelity 
\begin{eqnarray}
(1\!-\!F)   \ \sim \ 
\frac{9 \hbar^2 \abs{r_{eg}}^2}{v_0^2 \tau^4 \mcE_{eg}^2}
\label{eq:trap-IF}
\end{eqnarray}
where $F$ is defined as the fidelity. This $1-F$ factor plays a similar role to $K(t_f)$ in Eq.~(10) in the body of our letter; it determines both the deviations from ideal fidelity, and the entropy change of the qubit state.

Up to now, our derivation has been agnostic with respect to the shape of the trapping potential. However, the relation between the deviation we estimated above and the wavepacket spread $\Delta x$ depends on the exact shape of the trapping potential. Assuming that we are in a regime where the ideal dynamics is well-implemented even if not perfectly, it makes sense to assume that the relevant part of the potential around the minimum can be approximated as a harmonic potential. In this case, we have that $\abs{r_{eg}}^2 = \Delta x^2$ and $\mcE_{eg} = \hbar^2/(2 m \Delta x^2)$, where $\Delta x^2 = \ev{\hx^2}{g_K}= \hbar/(2 m \omega)$. Therefore, our estimate in terms of $\Delta x$ gives an infidelity
\begin{equation}
\textrm{Harmonic Potential Trap:}
\qquad (1\!-\!F) \sim
\frac{9 \hbar^2 \abs{r_{eg}}^2}{v_0^2 \tau^4 \mcE_{eg}^2}
=
\frac{36 m^2}{\hbar^2 v_0^2 \tau^4}\Delta x^6.
\label{Eq:nonideal-term-harmonic}
\end{equation}

Another simple model for which $\abs{r_{eg}}^2$ and $\mcE_{eg}$ can be easily computed is the infinite square well, or particle in a box. In this case, considering a box of length $L$ we have that $r_{eg} = 16 L/(9 \pi^2)$,  $\mcE_{eg} = 3 \hbar^2/(8 m L^2)$, and $\Delta x^2 = L^2 (\pi^2-6)/(12 \pi^2)$, i.e., $\Delta x \simeq 0.18 L$. It follows that then our estimate in terms of $\Delta x$ gives an infidelity
\begin{equation}
\textrm{Particle in a Box:}
\qquad (1\!-\!F) \sim
\frac{9 \hbar^2 \abs{r_{eg}}^2}{v_0^2 \tau^4 \mcE_{eg}^2}
=
\frac{2^{14} m^2}{3^4 \pi^4 \hbar^2 v_0^2 \tau^4} L^6
=
\frac{2^{20} \pi^2 m^2}{3 (\pi^2-6)^3\hbar^2 v_0^2 \tau^4} \Delta x^6
\sim
6\times 10^4 \frac{m^2}{\hbar^2 v_0^2 \tau^4} \Delta x^6.
\label{Eq:nonideal-term-square-well}
\end{equation}
We see that, while the scaling is again given by $\Delta x^6$, the deviation from ideal is three orders of magnitude higher that for a harmonic trap. This implies that a harmonic trap gives to an evolution closer to the ideal one than an infinite square well trap.

The fact that the $\Delta x$ scaling should be the same in Eqs.~(\ref{Eq:nonideal-term-harmonic}) and (\ref{Eq:nonideal-term-square-well}) can be seen from simple dimensional analysis.
Just looking at dimensions tells us that any trapping potential with a single lengthscale proportional to $\Delta x$ (such as harmonic potentials and infinite square wells) has $|r_{eg}|^2 \propto \Delta x^2$ and $\mcE_{eg} \propto \Delta x^{-2}$, hence Eq.~(\ref{eq:trap-IF}) tells us that it will exhibit decoherence that scales like $\Delta x^6$, although this will not give the all important prefactor in front of $\Delta x^6$.  In contrast, many trapping potentials have multiple lengthscales, such as a double-well potential with two lengthscales: the width of each well, and the distance between wells. In such cases, this simple dimensional analysis will not give the scaling with these various lengthscales.

\subsection{Typical magnitude of reservoir-free decoherence in surfing or shuttling experiments\label{APPSec:TrappingPotential_Magnitude}}

Above we showed that the dimensionless parameter determining deviations from ideal unitary dynamics is estimated to be smaller than $36 m^2(\Delta x)^6\big/\big(\hbar^2 v_0^2\tau^4\big)$ when considering harmonic potential traps.

Here we will extract numbers for a few experimental situations and show that 
$36 m^2(\Delta x)^6\big/\big(\hbar^2 v_0^2\tau^4\big)$ is completely negligible; less than $10^{-9}$.  
Thus the mechanism of reservoir-free decoherence discussed in this work will not be a significant source of imperfections for experimental qubits that fly by being trapped in the ground state of a moving classical potential. This is positive news, it means that such qubits are much more robust than ballistic qubits, and so are more suitable to process quantum information.\\

\emph{Surfing spin qubits:}
These qubits are electrons that are carried across a chip by a moving potential well created by a surface-acoustic wave.
The electron typically starts in a quantum dot, where quantum information can be inserted into its spin DoF. It is then released from the dot to be carried in the moving potential well to another quantum dot, where the information in its spin-state can be used.

Numbers vary by about an order of magnitude between experiments. Let us first take the following parameters, and then discuss variations: 
\begin{eqnarray}
\Delta x \ &\sim&\  10 \,{\rm nm}
\\
m \ &\sim&\ \mbox{electron mass} \ \sim\ 10^{-31}\,{\rm kg}
\\
v_0 \ &\sim&\ 10^4\,{\rm ms}^{-1}
\\
l\ &=&\ v_0\tau \ \sim\ 1\,{\rm \mu m}
\implies \tau \sim 10^{-10} {\rm s}.
\end{eqnarray}
Here $\Delta x$ and $v_0$ are taken from experiments,
but $l=v_0\tau$ is an estimate of the length over which the gate acts.

This gives 
\begin{eqnarray}
\frac{36 m^2(\Delta x)^6}{\hbar^2 v_0^2\tau^4} \ &\sim& 10^{-9}, 
\end{eqnarray}
which is small enough to neglect entirely. 
It will remain extremely small, even if we change multiple parameters by an order of magnitude.  
If an experimental system does enter a regime where this parameter is a bit too large for a given application, one can always reduce it strongly by making the trapping potential stronger to reduce $\Delta x$. A factor of two reduction in $\Delta x$ reduces the bound of deviations from ideal dynamics by a factor of 64. 

By comparison the fidelity of the current (first-generation) of surfing-qubit experiments is between 0.5 and 0.9.  Ref.~\cite{Takada2019Sound} provide numerics that shows that the principle cause of 
loss of fidelity is unrelated to the mechanism that we discuss here.
Instead it is that when the electron gets caught by the surface acoustic wave,
(rapidly accelerating the electron from stationary to 1000 ms$^{-1}$), it gets trapped in 
a superposition of highly excited states of the trapping potential, 
rather than getting trapped in the ground state.  Ref.~\cite{Najera-Santos2020Jul} shows how to control this effect, by making the trapping potential stronger. Our results here show that once this is done, there is nothing to stop surfing qubits having very high fidelities, much higher than
ballistic qubits.\\

\emph{Shuttling qubits:}
These qubits are extremely similar to the surfing qubits, 
the difference is how the trapping potential that moves is generated by time-dependent driving of gate-voltages, rather than a surface acoustic wave.
This results in a much smaller velocity for the moving trapping potential $v_0\sim 10\,{\rm ms}^{-1}$ and, with the same $l$ this implies a much bigger gate time: $\tau \sim 10^{-7} {\rm s}$.  If we assume that other parameters are similar to surfing qubits, we have
\begin{eqnarray}
36\frac{m^2(\Delta x)^6}{\hbar^2 v_0^2\tau^4} \ &\sim& 10^{-15}
\end{eqnarray}
Thus the reservoir-free decoherence effect will be absolutely negligible in such 
a system.\\

\emph{Flying Rydberg atoms:}
These are a hybrid case; the atom's center of mass flies ballistically, but the electronic states that carry the quantum information are confined by the trapping potential formed by the attractive potential of the nucleus.
At a handwaving level we can argue that the reservoir-free decoherence will be the sum of a term coming from the spread of wavepacket of the atom's center of mass, and a second term coming from the spread of the electronic wavepacket in the trapping potential generated by the attractive potential of the nucleus. Let us treat the second of these alone, by assuming that the nucleus is an infinite-mass point-like particle, generating a moving classical potential felt by the electronic states. If we approximate this as a harmonic trapping potential, we can estimate \cite{Najera-Santos2020Jul};
\begin{eqnarray}
\Delta x \ &\sim&\  10 \,{\rm \mu m}
\\
m \ &\sim&\ \mbox{electron mass} \ \sim\ 10^{-31}\,{\rm kg}
\\
v_0 \ &\sim&\ 10^4\,{\rm ms}^{-1}
\\
l\ &=&\ v_0\tau \ \sim\ 1\,{\rm cm}
\end{eqnarray}
This gives
\begin{eqnarray}
36\frac{m^2(\Delta x)^6}{\hbar^2 v_0^2\tau^4} \ &\sim& 10^{-7}
\end{eqnarray}
which is small enough to neglect entirely.
However, unlike in the case of surfing and shuttling qubits, 
the true trapping potential is not harmonic, so this estimate should be supplemented by a more careful calculation. This would be particularly important, if future experiments reduce the length-scale $l$ 
from a centimeter to a millimeter, for which this estimate would indicate that decoherence may cease to be negligible. 
As the atomic potential is given by nature, there is no easy way to change $\Delta x$ (unlike for surfing or shuttling spin-qubits), so there is no easy way to reduce the reservoir-free decoherence in flying atoms in situations where it becomes problematic.

\end{document}

%% file: preamble_packages.tex
\usepackage[T1]{fontenc}
\usepackage{graphicx}
\graphicspath{{./Figures/}}
\usepackage{dcolumn}
\usepackage{bm}
\usepackage[dvipsnames]{xcolor}
\usepackage{hyperref}
\hypersetup{colorlinks,linkcolor={MidnightBlue},citecolor={MidnightBlue},urlcolor={MidnightBlue}}  

\usepackage{float}

\usepackage{physics}
\usepackage{amsfonts}
\usepackage{verbatim}
\usepackage{amsmath}
\usepackage{csquotes}
\usepackage{color}
\usepackage{soul}
\usepackage{amsthm}
\usepackage{bm}
\usepackage{graphicx}
\usepackage[normalem]{ulem}

\usepackage{verbatim}


%% file: preamble_commands.tex
\newenvironment{equations}
{\begin{equation}\begin{aligned}}
		{\end{aligned}\end{equation}\ignorespacesafterend}

\newcommand{\prt}[1]{\left(#1\right)}

\newcommand{\Schr}{Schr\"{o}dinger\ }

\newcommand{\dg}{\dagger}

\newcommand{\prtq}[1]{\left[#1\right]}

\newcommand{\prtg}[1]{\left\{#1\right\}}

\newcommand{\prtqB}[1]{\Bigg[#1\Bigg]}

\newcommand{\prtgB}[1]{\Bigg\{#1\Bigg\}}

\newcommand{\mcE}{\mathcal{E}}

\newcommand{\mcT}{\mathcal{T}}

\newcommand{\mcU}{\mathcal{U}}

\newcommand{\mcK}{\mathcal{K}}

\newcommand{\mcC}{\mathcal{C}}

\newcommand{\Id}{\mathbb{I}}

\newcommand{\intmp}{\int_{-\infty}^{+\infty}}

\newcommand{\hc}{\textrm{h.c.}}

\newcommand{\hx}{\hat{x}}
\newcommand{\hp}{\hat{p}}
\newcommand{\hq}{\hat{q}}

\newcommand{\tm}{\tilde{m}}

\newcommand{\tH}{\widetilde{H}}

\newcommand{\tU}{\widetilde{U}}

\newcommand{\rC}{\rho}


%% file: LastArxivVersion.bbl
\begin{thebibliography}{53}%
	\makeatletter
	\providecommand \@ifxundefined [1]{%
		\@ifx{#1\undefined}
	}%
	\providecommand \@ifnum [1]{%
		\ifnum #1\expandafter \@firstoftwo
		\else \expandafter \@secondoftwo
		\fi
	}%
	\providecommand \@ifx [1]{%
		\ifx #1\expandafter \@firstoftwo
		\else \expandafter \@secondoftwo
		\fi
	}%
	\providecommand \natexlab [1]{#1}%
	\providecommand \enquote  [1]{``#1''}%
	\providecommand \bibnamefont  [1]{#1}%
	\providecommand \bibfnamefont [1]{#1}%
	\providecommand \citenamefont [1]{#1}%
	\providecommand \href@noop [0]{\@secondoftwo}%
	\providecommand \href [0]{\begingroup \@sanitize@url \@href}%
	\providecommand \@href[1]{\@@startlink{#1}\@@href}%
	\providecommand \@@href[1]{\endgroup#1\@@endlink}%
	\providecommand \@sanitize@url [0]{\catcode `\\12\catcode `\$12\catcode
		`\&12\catcode `\#12\catcode `\^12\catcode `\_12\catcode `\%12\relax}%
	\providecommand \@@startlink[1]{}%
	\providecommand \@@endlink[0]{}%
	\providecommand \url  [0]{\begingroup\@sanitize@url \@url }%
	\providecommand \@url [1]{\endgroup\@href {#1}{\urlprefix }}%
	\providecommand \urlprefix  [0]{URL }%
	\providecommand \Eprint [0]{\href }%
	\providecommand \doibase [0]{https://doi.org/}%
	\providecommand \selectlanguage [0]{\@gobble}%
	\providecommand \bibinfo  [0]{\@secondoftwo}%
	\providecommand \bibfield  [0]{\@secondoftwo}%
	\providecommand \translation [1]{[#1]}%
	\providecommand \BibitemOpen [0]{}%
	\providecommand \bibitemStop [0]{}%
	\providecommand \bibitemNoStop [0]{.\EOS\space}%
	\providecommand \EOS [0]{\spacefactor3000\relax}%
	\providecommand \BibitemShut  [1]{\csname bibitem#1\endcsname}%
	\let\auto@bib@innerbib\@empty
	\bibitem [{\citenamefont {Haroche}\ and\ \citenamefont
		{Raimond}(2006)}]{Book_Haroche2006}%
	\BibitemOpen
	\bibfield  {author} {\bibinfo {author} {\bibfnamefont {S.}~\bibnamefont
			{Haroche}}\ and\ \bibinfo {author} {\bibfnamefont {J.-M.}\ \bibnamefont
			{Raimond}},\ }\href
	{https://doi.org/10.1093/acprof:oso/9780198509141.001.0001} {\emph {\bibinfo
			{title} {Exploring The Quantum: Atoms, Cavities, and Photons}}}\ (\bibinfo
	{publisher} {Oxford University Press},\ \bibinfo {year} {2006})\BibitemShut
	{NoStop}%
	\bibitem [{\citenamefont {Haroche}\ \emph {et~al.}(2020)\citenamefont
		{Haroche}, \citenamefont {Brune},\ and\ \citenamefont
		{Raimond}}]{Haroche2020Mar}%
	\BibitemOpen
	\bibfield  {author} {\bibinfo {author} {\bibfnamefont {S.}~\bibnamefont
			{Haroche}}, \bibinfo {author} {\bibfnamefont {M.}~\bibnamefont {Brune}},\
		and\ \bibinfo {author} {\bibfnamefont {J.~M.}\ \bibnamefont {Raimond}},\
	}\bibfield  {title} {\bibinfo {title} {{From cavity to circuit quantum
				electrodynamics}},\ }\href {https://doi.org/10.1038/s41567-020-0812-1}
	{\bibfield  {journal} {\bibinfo  {journal} {Nat. Phys.}\ }\textbf {\bibinfo
			{volume} {16}},\ \bibinfo {pages} {243} (\bibinfo {year} {2020})}\BibitemShut
	{NoStop}%
	\bibitem [{\citenamefont {Osnaghi}\ \emph {et~al.}(2001)\citenamefont
		{Osnaghi}, \citenamefont {Bertet}, \citenamefont {Auffeves}, \citenamefont
		{Maioli}, \citenamefont {Brune}, \citenamefont {Raimond},\ and\ \citenamefont
		{Haroche}}]{chinese-collision}%
	\BibitemOpen
	\bibfield  {author} {\bibinfo {author} {\bibfnamefont {S.}~\bibnamefont
			{Osnaghi}}, \bibinfo {author} {\bibfnamefont {P.}~\bibnamefont {Bertet}},
		\bibinfo {author} {\bibfnamefont {A.}~\bibnamefont {Auffeves}}, \bibinfo
		{author} {\bibfnamefont {P.}~\bibnamefont {Maioli}}, \bibinfo {author}
		{\bibfnamefont {M.}~\bibnamefont {Brune}}, \bibinfo {author} {\bibfnamefont
			{J.~M.}\ \bibnamefont {Raimond}},\ and\ \bibinfo {author} {\bibfnamefont
			{S.}~\bibnamefont {Haroche}},\ }\bibfield  {title} {\bibinfo {title}
		{Coherent control of an atomic collision in a cavity},\ }\href
	{https://doi.org/10.1103/PhysRevLett.87.037902} {\bibfield  {journal}
		{\bibinfo  {journal} {Phys. Rev. Lett.}\ }\textbf {\bibinfo {volume} {87}},\
		\bibinfo {pages} {037902} (\bibinfo {year} {2001})}\BibitemShut {NoStop}%
	\bibitem [{\citenamefont {Rauschenbeutel}\ \emph {et~al.}(1999)\citenamefont
		{Rauschenbeutel}, \citenamefont {Nogues}, \citenamefont {Osnaghi},
		\citenamefont {Bertet}, \citenamefont {Brune}, \citenamefont {Raimond},\ and\
		\citenamefont {Haroche}}]{gate-gilles}%
	\BibitemOpen
	\bibfield  {author} {\bibinfo {author} {\bibfnamefont {A.}~\bibnamefont
			{Rauschenbeutel}}, \bibinfo {author} {\bibfnamefont {G.}~\bibnamefont
			{Nogues}}, \bibinfo {author} {\bibfnamefont {S.}~\bibnamefont {Osnaghi}},
		\bibinfo {author} {\bibfnamefont {P.}~\bibnamefont {Bertet}}, \bibinfo
		{author} {\bibfnamefont {M.}~\bibnamefont {Brune}}, \bibinfo {author}
		{\bibfnamefont {J.~M.}\ \bibnamefont {Raimond}},\ and\ \bibinfo {author}
		{\bibfnamefont {S.}~\bibnamefont {Haroche}},\ }\bibfield  {title} {\bibinfo
		{title} {Coherent operation of a tunable quantum phase gate in cavity qed},\
	}\href {https://doi.org/10.1103/PhysRevLett.83.5166} {\bibfield  {journal}
		{\bibinfo  {journal} {Phys. Rev. Lett.}\ }\textbf {\bibinfo {volume} {83}},\
		\bibinfo {pages} {5166} (\bibinfo {year} {1999})}\BibitemShut {NoStop}%
	\bibitem [{\citenamefont {Rauschenbeutel}\ \emph {et~al.}(2000)\citenamefont
		{Rauschenbeutel}, \citenamefont {Nogues}, \citenamefont {Osnaghi},
		\citenamefont {Bertet}, \citenamefont {Brune}, \citenamefont {Raimond},\ and\
		\citenamefont {Haroche}}]{arno-ghz}%
	\BibitemOpen
	\bibfield  {author} {\bibinfo {author} {\bibfnamefont {A.}~\bibnamefont
			{Rauschenbeutel}}, \bibinfo {author} {\bibfnamefont {G.}~\bibnamefont
			{Nogues}}, \bibinfo {author} {\bibfnamefont {S.}~\bibnamefont {Osnaghi}},
		\bibinfo {author} {\bibfnamefont {P.}~\bibnamefont {Bertet}}, \bibinfo
		{author} {\bibfnamefont {M.}~\bibnamefont {Brune}}, \bibinfo {author}
		{\bibfnamefont {J.-M.}\ \bibnamefont {Raimond}},\ and\ \bibinfo {author}
		{\bibfnamefont {S.}~\bibnamefont {Haroche}},\ }\bibfield  {title} {\bibinfo
		{title} {Step-by-step engineered multiparticle entanglement},\ }\href
	{https://doi.org/10.1126/science.288.5473.2024} {\bibfield  {journal}
		{\bibinfo  {journal} {Science}\ }\textbf {\bibinfo {volume} {288}},\ \bibinfo
		{pages} {2024} (\bibinfo {year} {2000})}\BibitemShut {NoStop}%
	\bibitem [{\citenamefont {Najera-Santos}\ \emph {et~al.}(2020)\citenamefont
		{Najera-Santos}, \citenamefont {Camati}, \citenamefont
		{M{\ifmmode\acute{e}\else\'{e}\fi}tillon}, \citenamefont {Brune},
		\citenamefont {Raimond}, \citenamefont
		{Auff{\ifmmode\grave{e}\else\`{e}\fi}ves},\ and\ \citenamefont
		{Dotsenko}}]{Najera-Santos2020Jul}%
	\BibitemOpen
	\bibfield  {author} {\bibinfo {author} {\bibfnamefont {B.-L.}\ \bibnamefont
			{Najera-Santos}}, \bibinfo {author} {\bibfnamefont {P.~A.}\ \bibnamefont
			{Camati}}, \bibinfo {author} {\bibfnamefont {V.}~\bibnamefont
			{M{\ifmmode\acute{e}\else\'{e}\fi}tillon}}, \bibinfo {author} {\bibfnamefont
			{M.}~\bibnamefont {Brune}}, \bibinfo {author} {\bibfnamefont {J.-M.}\
			\bibnamefont {Raimond}}, \bibinfo {author} {\bibfnamefont {A.}~\bibnamefont
			{Auff{\ifmmode\grave{e}\else\`{e}\fi}ves}},\ and\ \bibinfo {author}
		{\bibfnamefont {I.}~\bibnamefont {Dotsenko}},\ }\bibfield  {title} {\bibinfo
		{title} {{Autonomous Maxwell's demon in a cavity QED system}},\ }\href
	{https://doi.org/10.1103/PhysRevResearch.2.032025} {\bibfield  {journal}
		{\bibinfo  {journal} {Phys. Rev. Res.}\ }\textbf {\bibinfo {volume} {2}},\
		\bibinfo {pages} {032025(R)} (\bibinfo {year} {2020})}\BibitemShut {NoStop}%
	\bibitem [{\citenamefont {Jadot}\ \emph {et~al.}(2021)\citenamefont {Jadot},
		\citenamefont {Mortemousque}, \citenamefont {Chanrion}, \citenamefont
		{Thiney}, \citenamefont {Ludwig}, \citenamefont {Wieck}, \citenamefont
		{Urdampilleta}, \citenamefont {B{\"a}uerle},\ and\ \citenamefont
		{Meunier}}]{Jadot2021Distant}%
	\BibitemOpen
	\bibfield  {author} {\bibinfo {author} {\bibfnamefont {B.}~\bibnamefont
			{Jadot}}, \bibinfo {author} {\bibfnamefont {P.-A.}\ \bibnamefont
			{Mortemousque}}, \bibinfo {author} {\bibfnamefont {E.}~\bibnamefont
			{Chanrion}}, \bibinfo {author} {\bibfnamefont {V.}~\bibnamefont {Thiney}},
		\bibinfo {author} {\bibfnamefont {A.}~\bibnamefont {Ludwig}}, \bibinfo
		{author} {\bibfnamefont {A.~D.}\ \bibnamefont {Wieck}}, \bibinfo {author}
		{\bibfnamefont {M.}~\bibnamefont {Urdampilleta}}, \bibinfo {author}
		{\bibfnamefont {C.}~\bibnamefont {B{\"a}uerle}},\ and\ \bibinfo {author}
		{\bibfnamefont {T.}~\bibnamefont {Meunier}},\ }\bibfield  {title} {\bibinfo
		{title} {Distant spin entanglement via fast and coherent electron
			shuttling},\ }\href {https://www.nature.com/articles/s41565-021-00846-y}
	{\bibfield  {journal} {\bibinfo  {journal} {Nature Nanotechnology}\ }\textbf
		{\bibinfo {volume} {16}},\ \bibinfo {pages} {570} (\bibinfo {year}
		{2021})}\BibitemShut {NoStop}%
	\bibitem [{\citenamefont {Noiri}\ \emph {et~al.}(2022)\citenamefont {Noiri},
		\citenamefont {Takeda}, \citenamefont {Nakajima}, \citenamefont {Kobayashi},
		\citenamefont {Sammak}, \citenamefont {Scappucci},\ and\ \citenamefont
		{Tarucha}}]{Noiri2022Sep}%
	\BibitemOpen
	\bibfield  {author} {\bibinfo {author} {\bibfnamefont {A.}~\bibnamefont
			{Noiri}}, \bibinfo {author} {\bibfnamefont {K.}~\bibnamefont {Takeda}},
		\bibinfo {author} {\bibfnamefont {T.}~\bibnamefont {Nakajima}}, \bibinfo
		{author} {\bibfnamefont {T.}~\bibnamefont {Kobayashi}}, \bibinfo {author}
		{\bibfnamefont {A.}~\bibnamefont {Sammak}}, \bibinfo {author} {\bibfnamefont
			{G.}~\bibnamefont {Scappucci}},\ and\ \bibinfo {author} {\bibfnamefont
			{S.}~\bibnamefont {Tarucha}},\ }\bibfield  {title} {\bibinfo {title} {{A
				shuttling-based two-qubit logic gate for linking distant silicon quantum
				processors}},\ }\href {https://doi.org/10.1038/s41467-022-33453-z} {\bibfield
		{journal} {\bibinfo  {journal} {Nat. Commun.}\ }\textbf {\bibinfo {volume}
			{13}},\ \bibinfo {pages} {1} (\bibinfo {year} {2022})}\BibitemShut {NoStop}%
	\bibitem [{\citenamefont {Seidler}\ \emph {et~al.}(2022)\citenamefont
		{Seidler}, \citenamefont {Struck}, \citenamefont {Xue}, \citenamefont
		{Focke}, \citenamefont {Trellenkamp}, \citenamefont {Bluhm},\ and\
		\citenamefont {Schreiber}}]{Seidler2022Aug}%
	\BibitemOpen
	\bibfield  {author} {\bibinfo {author} {\bibfnamefont {I.}~\bibnamefont
			{Seidler}}, \bibinfo {author} {\bibfnamefont {T.}~\bibnamefont {Struck}},
		\bibinfo {author} {\bibfnamefont {R.}~\bibnamefont {Xue}}, \bibinfo {author}
		{\bibfnamefont {N.}~\bibnamefont {Focke}}, \bibinfo {author} {\bibfnamefont
			{S.}~\bibnamefont {Trellenkamp}}, \bibinfo {author} {\bibfnamefont
			{H.}~\bibnamefont {Bluhm}},\ and\ \bibinfo {author} {\bibfnamefont {L.~R.}\
			\bibnamefont {Schreiber}},\ }\bibfield  {title} {\bibinfo {title}
		{{Conveyor-mode single-electron shuttling in Si/SiGe for a scalable quantum
				computing architecture}},\ }\href
	{https://doi.org/10.1038/s41534-022-00615-2} {\bibfield  {journal} {\bibinfo
			{journal} {npj Quantum Inf.}\ }\textbf {\bibinfo {volume} {8}},\ \bibinfo
		{pages} {1} (\bibinfo {year} {2022})}\BibitemShut {NoStop}%
	\bibitem [{\citenamefont {Edlbauer}\ \emph {et~al.}(2022)\citenamefont
		{Edlbauer}, \citenamefont {Wang}, \citenamefont {Crozes}, \citenamefont
		{Perrier}, \citenamefont {Ouacel}, \citenamefont {Geffroy}, \citenamefont
		{Georgiou}, \citenamefont {Chatzikyriakou}, \citenamefont {Lacerda-Santos},
		\citenamefont {Waintal} \emph {et~al.}}]{Edlbauer2022semiconductor}%
	\BibitemOpen
	\bibfield  {author} {\bibinfo {author} {\bibfnamefont {H.}~\bibnamefont
			{Edlbauer}}, \bibinfo {author} {\bibfnamefont {J.}~\bibnamefont {Wang}},
		\bibinfo {author} {\bibfnamefont {T.}~\bibnamefont {Crozes}}, \bibinfo
		{author} {\bibfnamefont {P.}~\bibnamefont {Perrier}}, \bibinfo {author}
		{\bibfnamefont {S.}~\bibnamefont {Ouacel}}, \bibinfo {author} {\bibfnamefont
			{C.}~\bibnamefont {Geffroy}}, \bibinfo {author} {\bibfnamefont
			{G.}~\bibnamefont {Georgiou}}, \bibinfo {author} {\bibfnamefont
			{E.}~\bibnamefont {Chatzikyriakou}}, \bibinfo {author} {\bibfnamefont
			{A.}~\bibnamefont {Lacerda-Santos}}, \bibinfo {author} {\bibfnamefont
			{X.}~\bibnamefont {Waintal}}, \emph {et~al.},\ }\bibfield  {title} {\bibinfo
		{title} {Semiconductor-based electron flying qubits: review on recent
			progress accelerated by numerical modelling},\ }\href
	{https://epjqt.epj.org/articles/epjqt/abs/2022/01/40507_2022_Article_139/40507_2022_Article_139.html}
	{\bibfield  {journal} {\bibinfo  {journal} {EPJ Quantum Technology}\ }\textbf
		{\bibinfo {volume} {9}},\ \bibinfo {pages} {21} (\bibinfo {year}
		{2022})}\BibitemShut {NoStop}%
	\bibitem [{\citenamefont {Yamamoto}\ \emph {et~al.}(2012)\citenamefont
		{Yamamoto}, \citenamefont {Takada}, \citenamefont {B{\"a}uerle},
		\citenamefont {Watanabe}, \citenamefont {Wieck},\ and\ \citenamefont
		{Tarucha}}]{Yamamoto2012Electrical}%
	\BibitemOpen
	\bibfield  {author} {\bibinfo {author} {\bibfnamefont {M.}~\bibnamefont
			{Yamamoto}}, \bibinfo {author} {\bibfnamefont {S.}~\bibnamefont {Takada}},
		\bibinfo {author} {\bibfnamefont {C.}~\bibnamefont {B{\"a}uerle}}, \bibinfo
		{author} {\bibfnamefont {K.}~\bibnamefont {Watanabe}}, \bibinfo {author}
		{\bibfnamefont {A.~D.}\ \bibnamefont {Wieck}},\ and\ \bibinfo {author}
		{\bibfnamefont {S.}~\bibnamefont {Tarucha}},\ }\bibfield  {title} {\bibinfo
		{title} {Electrical control of a solid-state flying qubit},\ }\href
	{https://www.nature.com/articles/nnano.2012.28?WT} {\bibfield  {journal}
		{\bibinfo  {journal} {Nature Nanotechnology}\ }\textbf {\bibinfo {volume}
			{7}},\ \bibinfo {pages} {247} (\bibinfo {year} {2012})}\BibitemShut {NoStop}%
	\bibitem [{\citenamefont {B{\"a}uerle}\ \emph {et~al.}(2018)\citenamefont
		{B{\"a}uerle}, \citenamefont {Glattli}, \citenamefont {Meunier},
		\citenamefont {Portier}, \citenamefont {Roche}, \citenamefont {Roulleau},
		\citenamefont {Takada},\ and\ \citenamefont {Waintal}}]{Bauerle2018Coherent}%
	\BibitemOpen
	\bibfield  {author} {\bibinfo {author} {\bibfnamefont {C.}~\bibnamefont
			{B{\"a}uerle}}, \bibinfo {author} {\bibfnamefont {D.~C.}\ \bibnamefont
			{Glattli}}, \bibinfo {author} {\bibfnamefont {T.}~\bibnamefont {Meunier}},
		\bibinfo {author} {\bibfnamefont {F.}~\bibnamefont {Portier}}, \bibinfo
		{author} {\bibfnamefont {P.}~\bibnamefont {Roche}}, \bibinfo {author}
		{\bibfnamefont {P.}~\bibnamefont {Roulleau}}, \bibinfo {author}
		{\bibfnamefont {S.}~\bibnamefont {Takada}},\ and\ \bibinfo {author}
		{\bibfnamefont {X.}~\bibnamefont {Waintal}},\ }\bibfield  {title} {\bibinfo
		{title} {Coherent control of single electrons: a review of current
			progress},\ }\href
	{https://iopscience.iop.org/article/10.1088/1361-6633/aaa98a} {\bibfield
		{journal} {\bibinfo  {journal} {Reports on Progress in Physics}\ }\textbf
		{\bibinfo {volume} {81}},\ \bibinfo {pages} {056503} (\bibinfo {year}
		{2018})}\BibitemShut {NoStop}%
	\bibitem [{\citenamefont {Takada}\ \emph {et~al.}(2019)\citenamefont {Takada},
		\citenamefont {Edlbauer}, \citenamefont {Lepage}, \citenamefont {Wang},
		\citenamefont {Mortemousque}, \citenamefont {Georgiou}, \citenamefont
		{Barnes}, \citenamefont {Ford}, \citenamefont {Yuan}, \citenamefont {Santos}
		\emph {et~al.}}]{Takada2019Sound}%
	\BibitemOpen
	\bibfield  {author} {\bibinfo {author} {\bibfnamefont {S.}~\bibnamefont
			{Takada}}, \bibinfo {author} {\bibfnamefont {H.}~\bibnamefont {Edlbauer}},
		\bibinfo {author} {\bibfnamefont {H.~V.}\ \bibnamefont {Lepage}}, \bibinfo
		{author} {\bibfnamefont {J.}~\bibnamefont {Wang}}, \bibinfo {author}
		{\bibfnamefont {P.-A.}\ \bibnamefont {Mortemousque}}, \bibinfo {author}
		{\bibfnamefont {G.}~\bibnamefont {Georgiou}}, \bibinfo {author}
		{\bibfnamefont {C.~H.}\ \bibnamefont {Barnes}}, \bibinfo {author}
		{\bibfnamefont {C.~J.}\ \bibnamefont {Ford}}, \bibinfo {author}
		{\bibfnamefont {M.}~\bibnamefont {Yuan}}, \bibinfo {author} {\bibfnamefont
			{P.~V.}\ \bibnamefont {Santos}}, \emph {et~al.},\ }\bibfield  {title}
	{\bibinfo {title} {Sound-driven single-electron transfer in a circuit of
			coupled quantum rails},\ }\href
	{https://www.nature.com/articles/s41467-019-12514-w} {\bibfield  {journal}
		{\bibinfo  {journal} {Nature communications}\ }\textbf {\bibinfo {volume}
			{10}},\ \bibinfo {pages} {4557} (\bibinfo {year} {2019})}\BibitemShut
	{NoStop}%
	\bibitem [{\citenamefont {Freise}\ \emph {et~al.}(2020)\citenamefont {Freise},
		\citenamefont {Gerster}, \citenamefont {Reifert}, \citenamefont {Weimann},
		\citenamefont {Pierz}, \citenamefont {Hohls},\ and\ \citenamefont
		{Ubbelohde}}]{Freise2020Trapping}%
	\BibitemOpen
	\bibfield  {author} {\bibinfo {author} {\bibfnamefont {L.}~\bibnamefont
			{Freise}}, \bibinfo {author} {\bibfnamefont {T.}~\bibnamefont {Gerster}},
		\bibinfo {author} {\bibfnamefont {D.}~\bibnamefont {Reifert}}, \bibinfo
		{author} {\bibfnamefont {T.}~\bibnamefont {Weimann}}, \bibinfo {author}
		{\bibfnamefont {K.}~\bibnamefont {Pierz}}, \bibinfo {author} {\bibfnamefont
			{F.}~\bibnamefont {Hohls}},\ and\ \bibinfo {author} {\bibfnamefont
			{N.}~\bibnamefont {Ubbelohde}},\ }\bibfield  {title} {\bibinfo {title}
		{Trapping and counting ballistic nonequilibrium electrons},\ }\href
	{https://doi.org/10.1103/PhysRevLett.124.127701} {\bibfield  {journal}
		{\bibinfo  {journal} {Phys. Rev. Lett.}\ }\textbf {\bibinfo {volume} {124}},\
		\bibinfo {pages} {127701} (\bibinfo {year} {2020})}\BibitemShut {NoStop}%
	\bibitem [{\citenamefont {Edlbauer}\ \emph {et~al.}(2021)\citenamefont
		{Edlbauer}, \citenamefont {Wang}, \citenamefont {Ota}, \citenamefont
		{Richard}, \citenamefont {Jadot}, \citenamefont {Mortemousque}, \citenamefont
		{Okazaki}, \citenamefont {Nakamura}, \citenamefont {Kodera}, \citenamefont
		{Kaneko} \emph {et~al.}}]{Edlbauer2021Flight}%
	\BibitemOpen
	\bibfield  {author} {\bibinfo {author} {\bibfnamefont {H.}~\bibnamefont
			{Edlbauer}}, \bibinfo {author} {\bibfnamefont {J.}~\bibnamefont {Wang}},
		\bibinfo {author} {\bibfnamefont {S.}~\bibnamefont {Ota}}, \bibinfo {author}
		{\bibfnamefont {A.}~\bibnamefont {Richard}}, \bibinfo {author} {\bibfnamefont
			{B.}~\bibnamefont {Jadot}}, \bibinfo {author} {\bibfnamefont {P.-A.}\
			\bibnamefont {Mortemousque}}, \bibinfo {author} {\bibfnamefont
			{Y.}~\bibnamefont {Okazaki}}, \bibinfo {author} {\bibfnamefont
			{S.}~\bibnamefont {Nakamura}}, \bibinfo {author} {\bibfnamefont
			{T.}~\bibnamefont {Kodera}}, \bibinfo {author} {\bibfnamefont {N.-H.}\
			\bibnamefont {Kaneko}}, \emph {et~al.},\ }\bibfield  {title} {\bibinfo
		{title} {In-flight distribution of an electron within a surface acoustic
			wave},\ }\href {https://aip.scitation.org/doi/abs/10.1063/5.0062491}
	{\bibfield  {journal} {\bibinfo  {journal} {Applied Physics Letters}\
		}\textbf {\bibinfo {volume} {119}},\ \bibinfo {pages} {114004} (\bibinfo
		{year} {2021})}\BibitemShut {NoStop}%
	\bibitem [{\citenamefont {Wang}\ \emph {et~al.}(2023)\citenamefont {Wang},
		\citenamefont {Edlbauer}, \citenamefont {Richard}, \citenamefont {Ota},
		\citenamefont {Park}, \citenamefont {Shim}, \citenamefont {Ludwig},
		\citenamefont {Wieck}, \citenamefont {Sim}, \citenamefont {Urdampilleta},
		\citenamefont {Meunier}, \citenamefont {Kodera}, \citenamefont {Kaneko},
		\citenamefont {Sellier}, \citenamefont {Waintal}, \citenamefont {Takada},\
		and\ \citenamefont {B{\ifmmode\ddot{a}\else\"{a}\fi}uerle}}]{Wang2023Jul}%
	\BibitemOpen
	\bibfield  {author} {\bibinfo {author} {\bibfnamefont {J.}~\bibnamefont
			{Wang}}, \bibinfo {author} {\bibfnamefont {H.}~\bibnamefont {Edlbauer}},
		\bibinfo {author} {\bibfnamefont {A.}~\bibnamefont {Richard}}, \bibinfo
		{author} {\bibfnamefont {S.}~\bibnamefont {Ota}}, \bibinfo {author}
		{\bibfnamefont {W.}~\bibnamefont {Park}}, \bibinfo {author} {\bibfnamefont
			{J.}~\bibnamefont {Shim}}, \bibinfo {author} {\bibfnamefont {A.}~\bibnamefont
			{Ludwig}}, \bibinfo {author} {\bibfnamefont {A.~D.}\ \bibnamefont {Wieck}},
		\bibinfo {author} {\bibfnamefont {H.-S.}\ \bibnamefont {Sim}}, \bibinfo
		{author} {\bibfnamefont {M.}~\bibnamefont {Urdampilleta}}, \bibinfo {author}
		{\bibfnamefont {T.}~\bibnamefont {Meunier}}, \bibinfo {author} {\bibfnamefont
			{T.}~\bibnamefont {Kodera}}, \bibinfo {author} {\bibfnamefont {N.-H.}\
			\bibnamefont {Kaneko}}, \bibinfo {author} {\bibfnamefont {H.}~\bibnamefont
			{Sellier}}, \bibinfo {author} {\bibfnamefont {X.}~\bibnamefont {Waintal}},
		\bibinfo {author} {\bibfnamefont {S.}~\bibnamefont {Takada}},\ and\ \bibinfo
		{author} {\bibfnamefont {C.}~\bibnamefont
			{B{\ifmmode\ddot{a}\else\"{a}\fi}uerle}},\ }\bibfield  {title} {\bibinfo
		{title} {{Coulomb-mediated antibunching of an electron pair surfing on
				sound}},\ }\href {https://doi.org/10.1038/s41565-023-01368-5} {\bibfield
		{journal} {\bibinfo  {journal} {Nat. Nanotechnol.}\ }\textbf {\bibinfo
			{volume} {18}},\ \bibinfo {pages} {721} (\bibinfo {year} {2023})}\BibitemShut
	{NoStop}%
	\bibitem [{\citenamefont {Henkel}\ and\ \citenamefont
		{Folman}(2022)}]{Henkel2022Jun}%
	\BibitemOpen
	\bibfield  {author} {\bibinfo {author} {\bibfnamefont {C.}~\bibnamefont
			{Henkel}}\ and\ \bibinfo {author} {\bibfnamefont {R.}~\bibnamefont
			{Folman}},\ }\bibfield  {title} {\bibinfo {title} {{Internal decoherence in
				nano-object interferometry due to phonons}},\ }\href
	{https://doi.org/10.1116/5.0080503} {\bibfield  {journal} {\bibinfo
			{journal} {AVS Quantum Sci.}\ }\textbf {\bibinfo {volume} {4}},\ \bibinfo
		{pages} {025602} (\bibinfo {year} {2022})}\BibitemShut {NoStop}%
	\bibitem [{Sup()}]{note_autonomous}%
	\BibitemOpen
	\href@noop {} {\bibinfo {title} {In fields such as quantum thermodynamics, a system is described as autonomous if the Hamiltonian describing it is time independent.}}\BibitemShut {Stop}%
	\bibitem [{\citenamefont {Aharonov}\ and\ \citenamefont
		{Kaufherr}(1984)}]{Aharonov1984Quantum}%
	\BibitemOpen
	\bibfield  {author} {\bibinfo {author} {\bibfnamefont {Y.}~\bibnamefont
			{Aharonov}}\ and\ \bibinfo {author} {\bibfnamefont {T.}~\bibnamefont
			{Kaufherr}},\ }\bibfield  {title} {\bibinfo {title} {Quantum frames of
			reference},\ }\href {https://doi.org/10.1103/PhysRevD.30.368} {\bibfield
		{journal} {\bibinfo  {journal} {Phys. Rev. D}\ }\textbf {\bibinfo {volume}
			{30}},\ \bibinfo {pages} {368} (\bibinfo {year} {1984})}\BibitemShut
	{NoStop}%
	\bibitem [{\citenamefont {Aharonov}\ \emph {et~al.}(1998)\citenamefont
		{Aharonov}, \citenamefont {Oppenheim}, \citenamefont {Popescu}, \citenamefont
		{Reznik},\ and\ \citenamefont {Unruh}}]{Aharonov1998Measurement}%
	\BibitemOpen
	\bibfield  {author} {\bibinfo {author} {\bibfnamefont {Y.}~\bibnamefont
			{Aharonov}}, \bibinfo {author} {\bibfnamefont {J.}~\bibnamefont {Oppenheim}},
		\bibinfo {author} {\bibfnamefont {S.}~\bibnamefont {Popescu}}, \bibinfo
		{author} {\bibfnamefont {B.}~\bibnamefont {Reznik}},\ and\ \bibinfo {author}
		{\bibfnamefont {W.~G.}\ \bibnamefont {Unruh}},\ }\bibfield  {title} {\bibinfo
		{title} {Measurement of time of arrival in quantum mechanics},\ }\href
	{https://doi.org/10.1103/PhysRevA.57.4130} {\bibfield  {journal} {\bibinfo
			{journal} {Phys. Rev. A}\ }\textbf {\bibinfo {volume} {57}},\ \bibinfo
		{pages} {4130} (\bibinfo {year} {1998})}\BibitemShut {NoStop}%
	\bibitem [{\citenamefont {Gisin}\ and\ \citenamefont
		{Zambrini~Cruzeiro}(2018)}]{Gisin2018Quantum}%
	\BibitemOpen
	\bibfield  {author} {\bibinfo {author} {\bibfnamefont {N.}~\bibnamefont
			{Gisin}}\ and\ \bibinfo {author} {\bibfnamefont {E.}~\bibnamefont
			{Zambrini~Cruzeiro}},\ }\bibfield  {title} {\bibinfo {title} {Quantum
			measurements, energy conservation and quantum clocks},\ }\href
	{https://onlinelibrary.wiley.com/doi/10.1002/andp.201700388} {\bibfield
		{journal} {\bibinfo  {journal} {Annalen der Physik}\ }\textbf {\bibinfo
			{volume} {530}},\ \bibinfo {pages} {1700388} (\bibinfo {year}
		{2018})}\BibitemShut {NoStop}%
	\bibitem [{\citenamefont {Rogers}\ and\ \citenamefont
		{Jordan}(2022)}]{rogers2022postselection}%
	\BibitemOpen
	\bibfield  {author} {\bibinfo {author} {\bibfnamefont {S.}~\bibnamefont
			{Rogers}}\ and\ \bibinfo {author} {\bibfnamefont {A.~N.}\ \bibnamefont
			{Jordan}},\ }\bibfield  {title} {\bibinfo {title} {Postselection and quantum
			energetics},\ }\href@noop {} {\bibfield  {journal} {\bibinfo  {journal}
			{Physical Review A}\ }\textbf {\bibinfo {volume} {106}},\ \bibinfo {pages}
		{052214} (\bibinfo {year} {2022})}\BibitemShut {NoStop}%
	\bibitem [{\citenamefont {Jordan}\ and\ \citenamefont
		{Siddiqi}(2024)}]{jordan2024quantum}%
	\BibitemOpen
	\bibfield  {author} {\bibinfo {author} {\bibfnamefont {A.~N.}\ \bibnamefont
			{Jordan}}\ and\ \bibinfo {author} {\bibfnamefont {I.~A.}\ \bibnamefont
			{Siddiqi}},\ }\href@noop {} {\emph {\bibinfo {title} {Quantum Measurement:
				Theory and Practice}}}\ (\bibinfo  {publisher} {Cambridge University Press},\
	\bibinfo {year} {2024})\BibitemShut {NoStop}%
	\bibitem [{\citenamefont {Whitney}\ \emph {et~al.}(2008)\citenamefont
		{Whitney}, \citenamefont {Shnirman},\ and\ \citenamefont
		{Gefen}}]{Whitney2008Mar}%
	\BibitemOpen
	\bibfield  {author} {\bibinfo {author} {\bibfnamefont {R.~S.}\ \bibnamefont
			{Whitney}}, \bibinfo {author} {\bibfnamefont {A.}~\bibnamefont {Shnirman}},\
		and\ \bibinfo {author} {\bibfnamefont {Y.}~\bibnamefont {Gefen}},\ }\bibfield
	{title} {\bibinfo {title} {{Towards a Dephasing Diode: Asymmetric and
				Geometric Dephasing}},\ }\href
	{https://doi.org/10.1103/PhysRevLett.100.126806} {\bibfield  {journal}
		{\bibinfo  {journal} {Phys. Rev. Lett.}\ }\textbf {\bibinfo {volume} {100}},\
		\bibinfo {pages} {126806} (\bibinfo {year} {2008})}\BibitemShut {NoStop}%
	\bibitem [{\citenamefont {Englert}\ \emph {et~al.}(1991)\citenamefont
		{Englert}, \citenamefont {Schwinger}, \citenamefont {Barut},\ and\
		\citenamefont {Scully}}]{Englert1991reflecting}%
	\BibitemOpen
	\bibfield  {author} {\bibinfo {author} {\bibfnamefont {B.-G.}\ \bibnamefont
			{Englert}}, \bibinfo {author} {\bibfnamefont {J.}~\bibnamefont {Schwinger}},
		\bibinfo {author} {\bibfnamefont {A.}~\bibnamefont {Barut}},\ and\ \bibinfo
		{author} {\bibfnamefont {M.}~\bibnamefont {Scully}},\ }\bibfield  {title}
	{\bibinfo {title} {Reflecting slow atoms from a micromaser field},\ }\href
	{https://doi.org/10.1209/0295-5075/14/1/005} {\bibfield  {journal} {\bibinfo
			{journal} {EPL (Europhysics Letters)}\ }\textbf {\bibinfo {volume} {14}},\
		\bibinfo {pages} {25} (\bibinfo {year} {1991})}\BibitemShut {NoStop}%
	\bibitem [{\citenamefont {Scully}\ \emph {et~al.}(1996)\citenamefont {Scully},
		\citenamefont {Meyer},\ and\ \citenamefont {Walther}}]{Scully1996Induced}%
	\BibitemOpen
	\bibfield  {author} {\bibinfo {author} {\bibfnamefont {M.~O.}\ \bibnamefont
			{Scully}}, \bibinfo {author} {\bibfnamefont {G.~M.}\ \bibnamefont {Meyer}},\
		and\ \bibinfo {author} {\bibfnamefont {H.}~\bibnamefont {Walther}},\
	}\bibfield  {title} {\bibinfo {title} {Induced emission due to the quantized
			motion of ultracold atoms passing through a micromaser cavity},\ }\href
	{https://doi.org/10.1103/PhysRevLett.76.4144} {\bibfield  {journal} {\bibinfo
			{journal} {Phys. Rev. Lett.}\ }\textbf {\bibinfo {volume} {76}},\ \bibinfo
		{pages} {4144} (\bibinfo {year} {1996})}\BibitemShut {NoStop}%
	\bibitem [{\citenamefont {Larson}\ and\ \citenamefont
		{Abdel-Aty}(2009)}]{Larson2009Cavity}%
	\BibitemOpen
	\bibfield  {author} {\bibinfo {author} {\bibfnamefont {J.}~\bibnamefont
			{Larson}}\ and\ \bibinfo {author} {\bibfnamefont {M.}~\bibnamefont
			{Abdel-Aty}},\ }\bibfield  {title} {\bibinfo {title} {Cavity qed
			nondemolition measurement scheme using quantized atomic motion},\ }\href
	{https://doi.org/10.1103/PhysRevA.80.053609} {\bibfield  {journal} {\bibinfo
			{journal} {Phys. Rev. A}\ }\textbf {\bibinfo {volume} {80}},\ \bibinfo
		{pages} {053609} (\bibinfo {year} {2009})}\BibitemShut {NoStop}%
	\bibitem [{\citenamefont {Mercurio}\ \emph {et~al.}(2022)\citenamefont
		{Mercurio}, \citenamefont {De~Liberato}, \citenamefont {Nori}, \citenamefont
		{Savasta},\ and\ \citenamefont {Stassi}}]{Mercurio2022Flying}%
	\BibitemOpen
	\bibfield  {author} {\bibinfo {author} {\bibfnamefont {A.}~\bibnamefont
			{Mercurio}}, \bibinfo {author} {\bibfnamefont {S.}~\bibnamefont
			{De~Liberato}}, \bibinfo {author} {\bibfnamefont {F.}~\bibnamefont {Nori}},
		\bibinfo {author} {\bibfnamefont {S.}~\bibnamefont {Savasta}},\ and\ \bibinfo
		{author} {\bibfnamefont {R.}~\bibnamefont {Stassi}},\ }\bibfield  {title}
	{\bibinfo {title} {Flying atom back-reaction and mechanically generated
			photons from vacuum},\ }\href {https://arxiv.org/abs/2209.10419} {\bibfield
		{journal} {\bibinfo  {journal} {arXiv:2209.10419}\ } (\bibinfo {year}
		{2022})}\BibitemShut {NoStop}%
	\bibitem [{\citenamefont {Jacob}\ \emph {et~al.}(2021)\citenamefont {Jacob},
		\citenamefont {Esposito}, \citenamefont {Parrondo},\ and\ \citenamefont
		{Barra}}]{Jacob2021Thermalization}%
	\BibitemOpen
	\bibfield  {author} {\bibinfo {author} {\bibfnamefont {S.~L.}\ \bibnamefont
			{Jacob}}, \bibinfo {author} {\bibfnamefont {M.}~\bibnamefont {Esposito}},
		\bibinfo {author} {\bibfnamefont {J.~M.}\ \bibnamefont {Parrondo}},\ and\
		\bibinfo {author} {\bibfnamefont {F.}~\bibnamefont {Barra}},\ }\bibfield
	{title} {\bibinfo {title} {Thermalization induced by quantum scattering},\
	}\href {https://doi.org/10.1103/PRXQuantum.2.020312} {\bibfield  {journal}
		{\bibinfo  {journal} {PRX Quantum}\ }\textbf {\bibinfo {volume} {2}},\
		\bibinfo {pages} {020312} (\bibinfo {year} {2021})}\BibitemShut {NoStop}%
	\bibitem [{\citenamefont {Ciccarello}\ \emph {et~al.}(2022)\citenamefont
		{Ciccarello}, \citenamefont {Lorenzo}, \citenamefont {Giovannetti},\ and\
		\citenamefont {Palma}}]{Ciccarello2021Quantum}%
	\BibitemOpen
	\bibfield  {author} {\bibinfo {author} {\bibfnamefont {F.}~\bibnamefont
			{Ciccarello}}, \bibinfo {author} {\bibfnamefont {S.}~\bibnamefont {Lorenzo}},
		\bibinfo {author} {\bibfnamefont {V.}~\bibnamefont {Giovannetti}},\ and\
		\bibinfo {author} {\bibfnamefont {G.~M.}\ \bibnamefont {Palma}},\ }\bibfield
	{title} {\bibinfo {title} {Quantum collision models: Open system dynamics
			from repeated interactions},\ }\href
	{https://doi.org/https://doi.org/10.1016/j.physrep.2022.01.001} {\bibfield
		{journal} {\bibinfo  {journal} {Physics Reports}\ }\textbf {\bibinfo {volume}
			{954}},\ \bibinfo {pages} {1} (\bibinfo {year} {2022})}\BibitemShut {NoStop}%
	\bibitem [{\citenamefont {Jacob}\ \emph {et~al.}(2022)\citenamefont {Jacob},
		\citenamefont {Esposito}, \citenamefont {Parrondo},\ and\ \citenamefont
		{Barra}}]{Jacob2022QuantumScattering}%
	\BibitemOpen
	\bibfield  {author} {\bibinfo {author} {\bibfnamefont {S.~L.}\ \bibnamefont
			{Jacob}}, \bibinfo {author} {\bibfnamefont {M.}~\bibnamefont {Esposito}},
		\bibinfo {author} {\bibfnamefont {J.~M.~R.}\ \bibnamefont {Parrondo}},\ and\
		\bibinfo {author} {\bibfnamefont {F.}~\bibnamefont {Barra}},\ }\bibfield
	{title} {\bibinfo {title} {Quantum scattering as a work source},\ }\href
	{https://doi.org/10.22331/q-2022-06-29-750} {\bibfield  {journal} {\bibinfo
			{journal} {{Quantum}}\ }\textbf {\bibinfo {volume} {6}},\ \bibinfo {pages}
		{750} (\bibinfo {year} {2022})}\BibitemShut {NoStop}%
	\bibitem [{\citenamefont {Tabanera}\ \emph {et~al.}(2022)\citenamefont
		{Tabanera}, \citenamefont {Luque}, \citenamefont {Jacob}, \citenamefont
		{Esposito}, \citenamefont {Barra},\ and\ \citenamefont
		{Parrondo}}]{Tabanera2022Quantum}%
	\BibitemOpen
	\bibfield  {author} {\bibinfo {author} {\bibfnamefont {J.}~\bibnamefont
			{Tabanera}}, \bibinfo {author} {\bibfnamefont {I.}~\bibnamefont {Luque}},
		\bibinfo {author} {\bibfnamefont {S.~L.}\ \bibnamefont {Jacob}}, \bibinfo
		{author} {\bibfnamefont {M.}~\bibnamefont {Esposito}}, \bibinfo {author}
		{\bibfnamefont {F.}~\bibnamefont {Barra}},\ and\ \bibinfo {author}
		{\bibfnamefont {J.~M.~R.}\ \bibnamefont {Parrondo}},\ }\bibfield  {title}
	{\bibinfo {title} {Quantum collisional thermostats},\ }\href
	{https://doi.org/10.1088/1367-2630/ac4923} {\bibfield  {journal} {\bibinfo
			{journal} {New Journal of Physics}\ }\textbf {\bibinfo {volume} {24}},\
		\bibinfo {pages} {023018} (\bibinfo {year} {2022})}\BibitemShut {NoStop}%
	\bibitem [{\citenamefont {Binder}\ \emph {et~al.}(2018)\citenamefont {Binder},
		\citenamefont {Correa}, \citenamefont {Gogolin}, \citenamefont {Anders},\
		and\ \citenamefont {Adesso}}]{Book_Binder2018}%
	\BibitemOpen
	\bibfield  {author} {\bibinfo {author} {\bibfnamefont {F.}~\bibnamefont
			{Binder}}, \bibinfo {author} {\bibfnamefont {L.}~\bibnamefont {Correa}},
		\bibinfo {author} {\bibfnamefont {C.}~\bibnamefont {Gogolin}}, \bibinfo
		{author} {\bibfnamefont {J.}~\bibnamefont {Anders}},\ and\ \bibinfo {author}
		{\bibfnamefont {G.}~\bibnamefont {Adesso}},\ }\href
	{https://doi.org/10.1007/978-3-319-99046-0} {\emph {\bibinfo {title}
			{Thermodynamics in the Quantum Regime}}}\ (\bibinfo  {publisher} {Springer
		International Publishing},\ \bibinfo {year} {2018})\BibitemShut {NoStop}%
	\bibitem [{\citenamefont {Chiara}\ \emph {et~al.}(2018)\citenamefont {Chiara},
		\citenamefont {Landi}, \citenamefont {Hewgill}, \citenamefont {Reid},
		\citenamefont {Ferraro}, \citenamefont {Roncaglia},\ and\ \citenamefont
		{Antezza}}]{DeChiara2018}%
	\BibitemOpen
	\bibfield  {author} {\bibinfo {author} {\bibfnamefont {G.~D.}\ \bibnamefont
			{Chiara}}, \bibinfo {author} {\bibfnamefont {G.}~\bibnamefont {Landi}},
		\bibinfo {author} {\bibfnamefont {A.}~\bibnamefont {Hewgill}}, \bibinfo
		{author} {\bibfnamefont {B.}~\bibnamefont {Reid}}, \bibinfo {author}
		{\bibfnamefont {A.}~\bibnamefont {Ferraro}}, \bibinfo {author} {\bibfnamefont
			{A.~J.}\ \bibnamefont {Roncaglia}},\ and\ \bibinfo {author} {\bibfnamefont
			{M.}~\bibnamefont {Antezza}},\ }\bibfield  {title} {\bibinfo {title}
		{Reconciliation of quantum local master equations with thermodynamics},\
	}\href {https://doi.org/10.1088/1367-2630/aaecee} {\bibfield  {journal}
		{\bibinfo  {journal} {New J. Phys.}\ }\textbf {\bibinfo {volume} {20}},\
		\bibinfo {pages} {113024} (\bibinfo {year} {2018})}\BibitemShut {NoStop}%
	\bibitem [{\citenamefont {Los}\ and\ \citenamefont {Los}(2013)}]{Los2013Exact}%
	\BibitemOpen
	\bibfield  {author} {\bibinfo {author} {\bibfnamefont {V.}~\bibnamefont
			{Los}}\ and\ \bibinfo {author} {\bibfnamefont {N.}~\bibnamefont {Los}},\
	}\bibfield  {title} {\bibinfo {title} {Exact solution of the one-dimensional
			time-dependent schr{\"o}dinger equation with a rectangular well/barrier
			potential and its applications},\ }\href
	{https://link.springer.com/article/10.1007%2Fs11232-013-0128-8} {\bibfield
		{journal} {\bibinfo  {journal} {Theoretical and Mathematical Physics}\
		}\textbf {\bibinfo {volume} {177}},\ \bibinfo {pages} {1706} (\bibinfo {year}
		{2013})}\BibitemShut {NoStop}%
	\bibitem [{\citenamefont {Riahi}(2017)}]{Riahi2017Solving}%
	\BibitemOpen
	\bibfield  {author} {\bibinfo {author} {\bibfnamefont {N.}~\bibnamefont
			{Riahi}},\ }\bibfield  {title} {\bibinfo {title} {Solving the time-dependent
			schr{\"o}dinger equation via laplace transform},\ }\href
	{https://link.springer.com/article/10.1007/s40509-016-0087-5} {\bibfield
		{journal} {\bibinfo  {journal} {Quantum Studies: Mathematics and
				Foundations}\ }\textbf {\bibinfo {volume} {4}},\ \bibinfo {pages} {103}
		(\bibinfo {year} {2017})}\BibitemShut {NoStop}%
	\bibitem [{\citenamefont {Malabarba}\ \emph {et~al.}(2015)\citenamefont
		{Malabarba}, \citenamefont {Short},\ and\ \citenamefont
		{Kammerlander}}]{Malabarba2015clock}%
	\BibitemOpen
	\bibfield  {author} {\bibinfo {author} {\bibfnamefont {A.~S.}\ \bibnamefont
			{Malabarba}}, \bibinfo {author} {\bibfnamefont {A.~J.}\ \bibnamefont
			{Short}},\ and\ \bibinfo {author} {\bibfnamefont {P.}~\bibnamefont
			{Kammerlander}},\ }\bibfield  {title} {\bibinfo {title} {Clock-driven quantum
			thermal engines},\ }\href {https://doi.org/10.1088/1367-2630/17/4/045027}
	{\bibfield  {journal} {\bibinfo  {journal} {New Journal of Physics}\ }\textbf
		{\bibinfo {volume} {17}},\ \bibinfo {pages} {045027} (\bibinfo {year}
		{2015})}\BibitemShut {NoStop}%
	\bibitem [{\citenamefont {So\l{}tan}\ \emph {et~al.}(2021)\citenamefont
		{So\l{}tan}, \citenamefont {Fr\k{a}czak}, \citenamefont {Belzig},\ and\
		\citenamefont {Bednorz}}]{Soltan2021Conservation}%
	\BibitemOpen
	\bibfield  {author} {\bibinfo {author} {\bibfnamefont {S.}~\bibnamefont
			{So\l{}tan}}, \bibinfo {author} {\bibfnamefont {M.}~\bibnamefont
			{Fr\k{a}czak}}, \bibinfo {author} {\bibfnamefont {W.}~\bibnamefont
			{Belzig}},\ and\ \bibinfo {author} {\bibfnamefont {A.}~\bibnamefont
			{Bednorz}},\ }\bibfield  {title} {\bibinfo {title} {Conservation laws in
			quantum noninvasive measurements},\ }\href
	{https://doi.org/10.1103/PhysRevResearch.3.013247} {\bibfield  {journal}
		{\bibinfo  {journal} {Phys. Rev. Research}\ }\textbf {\bibinfo {volume}
			{3}},\ \bibinfo {pages} {013247} (\bibinfo {year} {2021})}\BibitemShut
	{NoStop}%
	\bibitem [{\citenamefont {Bresque}(2022)}]{Bresque2022energetique}%
	\BibitemOpen
	\bibfield  {author} {\bibinfo {author} {\bibfnamefont {L.}~\bibnamefont
			{Bresque}},\ }\emph {\bibinfo {title} {{\'E}nerg{\'e}tique de la mesure
			quantique}},\ \href {https://www.theses.fr/2022GRALY061} {Ph.D. thesis},\
	\bibinfo  {school} {Universit{\'e} Grenoble Alpes} (\bibinfo {year}
	{2022})\BibitemShut {NoStop}%
	\bibitem [{\citenamefont {Breuer}\ and\ \citenamefont
		{Petruccione}(2007)}]{Book_Breuer2002}%
	\BibitemOpen
	\bibfield  {author} {\bibinfo {author} {\bibfnamefont {H.-P.}\ \bibnamefont
			{Breuer}}\ and\ \bibinfo {author} {\bibfnamefont {F.}~\bibnamefont
			{Petruccione}},\ }\href
	{https://doi.org/10.1093/acprof:oso/9780199213900.001.0001} {\emph {\bibinfo
			{title} {The Theory of Open quantum systems}}}\ (\bibinfo  {publisher}
	{Oxford University Press, New York},\ \bibinfo {year} {2007})\BibitemShut
	{NoStop}%
	\bibitem [{\citenamefont {Nielsen}\ and\ \citenamefont
		{Chuang}(2010)}]{Book_Nielsen2010}%
	\BibitemOpen
	\bibfield  {author} {\bibinfo {author} {\bibfnamefont {M.~A.}\ \bibnamefont
			{Nielsen}}\ and\ \bibinfo {author} {\bibfnamefont {I.~L.}\ \bibnamefont
			{Chuang}},\ }\href {https://doi.org/10.1017/CBO9780511976667} {\emph
		{\bibinfo {title} {Quantum Computation and Quantum Information: 10th
				Anniversary Edition}}}\ (\bibinfo  {publisher} {Cambridge University Press},\
	\bibinfo {year} {2010})\BibitemShut {NoStop}%
	\bibitem [{\citenamefont {Grace}\ and\ \citenamefont
		{Guha}(2021)}]{Grace2021Perturbation}%
	\BibitemOpen
	\bibfield  {author} {\bibinfo {author} {\bibfnamefont {M.~R.}\ \bibnamefont
			{Grace}}\ and\ \bibinfo {author} {\bibfnamefont {S.}~\bibnamefont {Guha}},\
	}\bibfield  {title} {\bibinfo {title} {Perturbation theory for quantum
			information},\ }\href {https://arxiv.org/abs/2106.05533v1} {\bibfield
		{journal} {\bibinfo  {journal} {arXiv:2106.05533}\ } (\bibinfo {year}
		{2021})}\BibitemShut {NoStop}%
	\bibitem [{\citenamefont {Das}\ \emph {et~al.}(2018)\citenamefont {Das},
		\citenamefont {Khatri}, \citenamefont {Siopsis},\ and\ \citenamefont
		{Wilde}}]{Das2018Fundamental}%
	\BibitemOpen
	\bibfield  {author} {\bibinfo {author} {\bibfnamefont {S.}~\bibnamefont
			{Das}}, \bibinfo {author} {\bibfnamefont {S.}~\bibnamefont {Khatri}},
		\bibinfo {author} {\bibfnamefont {G.}~\bibnamefont {Siopsis}},\ and\ \bibinfo
		{author} {\bibfnamefont {M.~M.}\ \bibnamefont {Wilde}},\ }\bibfield  {title}
	{\bibinfo {title} {Fundamental limits on quantum dynamics based on entropy
			change},\ }\href {https://doi.org/10.1063/1.4997044} {\bibfield  {journal}
		{\bibinfo  {journal} {Journal of Mathematical Physics}\ }\textbf {\bibinfo
			{volume} {59}},\ \bibinfo {pages} {012205} (\bibinfo {year}
		{2018})}\BibitemShut {NoStop}%
	\bibitem [{\citenamefont {Schmied}(2020)}]{Book_Schmied2020Mathematica}%
	\BibitemOpen
	\bibfield  {author} {\bibinfo {author} {\bibfnamefont {R.}~\bibnamefont
			{Schmied}},\ }\href
	{https://link.springer.com/book/10.1007/978-981-13-7588-0} {\emph {\bibinfo
			{title} {Using Mathematica for Quantum Mechanics}}}\ (\bibinfo  {publisher}
	{Springer, Singapore},\ \bibinfo {year} {2020})\BibitemShut {NoStop}%
	\bibitem [{\citenamefont {Dubois}\ \emph {et~al.}(2013)\citenamefont {Dubois},
		\citenamefont {Jullien}, \citenamefont {Portier}, \citenamefont {Roche},
		\citenamefont {Cavanna}, \citenamefont {Jin}, \citenamefont {Wegscheider},
		\citenamefont {Roulleau},\ and\ \citenamefont {Glattli}}]{Dubois2013Oct}%
	\BibitemOpen
	\bibfield  {author} {\bibinfo {author} {\bibfnamefont {J.}~\bibnamefont
			{Dubois}}, \bibinfo {author} {\bibfnamefont {T.}~\bibnamefont {Jullien}},
		\bibinfo {author} {\bibfnamefont {F.}~\bibnamefont {Portier}}, \bibinfo
		{author} {\bibfnamefont {P.}~\bibnamefont {Roche}}, \bibinfo {author}
		{\bibfnamefont {A.}~\bibnamefont {Cavanna}}, \bibinfo {author} {\bibfnamefont
			{Y.}~\bibnamefont {Jin}}, \bibinfo {author} {\bibfnamefont {W.}~\bibnamefont
			{Wegscheider}}, \bibinfo {author} {\bibfnamefont {P.}~\bibnamefont
			{Roulleau}},\ and\ \bibinfo {author} {\bibfnamefont {D.~C.}\ \bibnamefont
			{Glattli}},\ }\bibfield  {title} {\bibinfo {title} {{Minimal-excitation
				states for electron quantum optics using levitons}},\ }\href
	{https://doi.org/10.1038/nature12713} {\bibfield  {journal} {\bibinfo
			{journal} {Nature}\ }\textbf {\bibinfo {volume} {502}},\ \bibinfo {pages}
		{659} (\bibinfo {year} {2013})}\BibitemShut {NoStop}%
	\bibitem [{\citenamefont {Freulon}\ \emph {et~al.}(2015)\citenamefont
		{Freulon}, \citenamefont {Marguerite}, \citenamefont {Berroir}, \citenamefont
		{Pla{\ifmmode\mbox{\c{c}}\else\c{c}\fi}ais}, \citenamefont {Cavanna},
		\citenamefont {Jin},\ and\ \citenamefont
		{F{\ifmmode\grave{e}\else\`{e}\fi}ve}}]{Freulon2015Apr}%
	\BibitemOpen
	\bibfield  {author} {\bibinfo {author} {\bibfnamefont {V.}~\bibnamefont
			{Freulon}}, \bibinfo {author} {\bibfnamefont {A.}~\bibnamefont {Marguerite}},
		\bibinfo {author} {\bibfnamefont {J.-M.}\ \bibnamefont {Berroir}}, \bibinfo
		{author} {\bibfnamefont {B.}~\bibnamefont
			{Pla{\ifmmode\mbox{\c{c}}\else\c{c}\fi}ais}}, \bibinfo {author}
		{\bibfnamefont {A.}~\bibnamefont {Cavanna}}, \bibinfo {author} {\bibfnamefont
			{Y.}~\bibnamefont {Jin}},\ and\ \bibinfo {author} {\bibfnamefont
			{G.}~\bibnamefont {F{\ifmmode\grave{e}\else\`{e}\fi}ve}},\ }\bibfield
	{title} {\bibinfo {title} {{Hong-Ou-Mandel experiment for temporal
				investigation of single-electron fractionalization}},\ }\href
	{https://doi.org/10.1038/ncomms7854} {\bibfield  {journal} {\bibinfo
			{journal} {Nat. Commun.}\ }\textbf {\bibinfo {volume} {6}},\ \bibinfo {pages}
		{1} (\bibinfo {year} {2015})}\BibitemShut {NoStop}%
	\bibitem [{\citenamefont {Kataoka}\ \emph {et~al.}(2016)\citenamefont
		{Kataoka}, \citenamefont {Johnson}, \citenamefont {Emary}, \citenamefont
		{See}, \citenamefont {Griffiths}, \citenamefont {Jones}, \citenamefont
		{Farrer}, \citenamefont {Ritchie}, \citenamefont {Pepper},\ and\
		\citenamefont {Janssen}}]{Kataoka2016Mar}%
	\BibitemOpen
	\bibfield  {author} {\bibinfo {author} {\bibfnamefont {M.}~\bibnamefont
			{Kataoka}}, \bibinfo {author} {\bibfnamefont {N.}~\bibnamefont {Johnson}},
		\bibinfo {author} {\bibfnamefont {C.}~\bibnamefont {Emary}}, \bibinfo
		{author} {\bibfnamefont {P.}~\bibnamefont {See}}, \bibinfo {author}
		{\bibfnamefont {J.~P.}\ \bibnamefont {Griffiths}}, \bibinfo {author}
		{\bibfnamefont {G.~A.~C.}\ \bibnamefont {Jones}}, \bibinfo {author}
		{\bibfnamefont {I.}~\bibnamefont {Farrer}}, \bibinfo {author} {\bibfnamefont
			{D.~A.}\ \bibnamefont {Ritchie}}, \bibinfo {author} {\bibfnamefont
			{M.}~\bibnamefont {Pepper}},\ and\ \bibinfo {author} {\bibfnamefont {T.~J.
				B.~M.}\ \bibnamefont {Janssen}},\ }\bibfield  {title} {\bibinfo {title}
		{{Time-of-Flight Measurements of Single-Electron Wave Packets in Quantum Hall
				Edge States}},\ }\href {https://doi.org/10.1103/PhysRevLett.116.126803}
	{\bibfield  {journal} {\bibinfo  {journal} {Phys. Rev. Lett.}\ }\textbf
		{\bibinfo {volume} {116}},\ \bibinfo {pages} {126803} (\bibinfo {year}
		{2016})}\BibitemShut {NoStop}%
	\bibitem [{\citenamefont {Ferraro}\ \emph {et~al.}(2018)\citenamefont
		{Ferraro}, \citenamefont {Ronetti}, \citenamefont {Vannucci}, \citenamefont
		{Acciai}, \citenamefont {Rech}, \citenamefont {Jockheere}, \citenamefont
		{Martin},\ and\ \citenamefont {Sassetti}}]{Ferraro2018Dec}%
	\BibitemOpen
	\bibfield  {author} {\bibinfo {author} {\bibfnamefont {D.}~\bibnamefont
			{Ferraro}}, \bibinfo {author} {\bibfnamefont {F.}~\bibnamefont {Ronetti}},
		\bibinfo {author} {\bibfnamefont {L.}~\bibnamefont {Vannucci}}, \bibinfo
		{author} {\bibfnamefont {M.}~\bibnamefont {Acciai}}, \bibinfo {author}
		{\bibfnamefont {J.}~\bibnamefont {Rech}}, \bibinfo {author} {\bibfnamefont
			{T.}~\bibnamefont {Jockheere}}, \bibinfo {author} {\bibfnamefont
			{T.}~\bibnamefont {Martin}},\ and\ \bibinfo {author} {\bibfnamefont
			{M.}~\bibnamefont {Sassetti}},\ }\bibfield  {title} {\bibinfo {title}
		{{Hong-Ou-Mandel characterization of multiply charged Levitons}},\ }\href
	{https://doi.org/10.1140/epjst/e2018-800074-1} {\bibfield  {journal}
		{\bibinfo  {journal} {Eur. Phys. J. Spec. Top.}\ }\textbf {\bibinfo {volume}
			{227}},\ \bibinfo {pages} {1345} (\bibinfo {year} {2018})}\BibitemShut
	{NoStop}%
	\bibitem [{Sup()}]{note_internal_state_obs}
	{Any internal-state observable can be measured by performing a unitary rotation followed by a measurement of populations in the energy eigenbasis of $H_0$.}\BibitemShut {Stop}%
	\bibitem [{Sup()}]{note_regime_harmonic}
	{Here, we are interested in a regime where the nonautonomous dynamics is well-but-not-perfectly implemented, just as we were for ballistic qubits earlier. As such, the choice of harmonic trap potential is of general interest, as many potentials behave harmonically around their minima for well-trapped particles.}\BibitemShut {Stop}%
	\bibitem [{Sup()}]{scaling}
	{This $\Delta x^6$ scaling assumes a harmonic trapping potential. While different $\Delta x$ scaling occurs for less usual shapes of trapping potentials, our SM [39] shows it to be $\Delta x^6$ for anypotential characterized by a single length scale $\Delta x$. It shows that an infinite square well also has $\Delta x^6$ scaling, but with a prefactor that is 3 orders of magnitude larger than for the harmonic potential.}\BibitemShut {Stop}%
	\bibitem [{\citenamefont {Weimer}\ \emph {et~al.}(2008)\citenamefont {Weimer},
		\citenamefont {Henrich}, \citenamefont {Rempp}, \citenamefont {Schröder},\
		and\ \citenamefont {Mahler}}]{Weimer2008Local}%
	\BibitemOpen
	\bibfield  {author} {\bibinfo {author} {\bibfnamefont {H.}~\bibnamefont
			{Weimer}}, \bibinfo {author} {\bibfnamefont {M.~J.}\ \bibnamefont {Henrich}},
		\bibinfo {author} {\bibfnamefont {F.}~\bibnamefont {Rempp}}, \bibinfo
		{author} {\bibfnamefont {H.}~\bibnamefont {Schröder}},\ and\ \bibinfo
		{author} {\bibfnamefont {G.}~\bibnamefont {Mahler}},\ }\bibfield  {title}
	{\bibinfo {title} {Local effective dynamics of quantum systems: A generalized
			approach to work and heat},\ }\href
	{https://doi.org/10.1209/0295-5075/83/30008} {\bibfield  {journal} {\bibinfo
			{journal} {{EPL} (Europhysics Letters)}\ }\textbf {\bibinfo {volume} {83}},\
		\bibinfo {pages} {30008} (\bibinfo {year} {2008})}\BibitemShut {NoStop}%
	\bibitem [{\citenamefont {Hossein-Nejad}\ \emph {et~al.}(2015)\citenamefont
		{Hossein-Nejad}, \citenamefont {O'Reilly},\ and\ \citenamefont
		{Olaya-Castro}}]{HosseinNejad2015Work}%
	\BibitemOpen
	\bibfield  {author} {\bibinfo {author} {\bibfnamefont {H.}~\bibnamefont
			{Hossein-Nejad}}, \bibinfo {author} {\bibfnamefont {E.~J.}\ \bibnamefont
			{O'Reilly}},\ and\ \bibinfo {author} {\bibfnamefont {A.}~\bibnamefont
			{Olaya-Castro}},\ }\bibfield  {title} {\bibinfo {title} {Work, heat and
			entropy production in bipartite quantum systems},\ }\href
	{https://doi.org/10.1088/1367-2630/17/7/075014} {\bibfield  {journal}
		{\bibinfo  {journal} {New Journal of Physics}\ }\textbf {\bibinfo {volume}
			{17}},\ \bibinfo {pages} {075014} (\bibinfo {year} {2015})}\BibitemShut
	{NoStop}%
	\bibitem [{\citenamefont {Xiao-Yu}(2010)}]{XiaoYu2010Perturbation}%
	\BibitemOpen
	\bibfield  {author} {\bibinfo {author} {\bibfnamefont {C.}~\bibnamefont
			{Xiao-Yu}},\ }\bibfield  {title} {\bibinfo {title} {Perturbation theory of
			von neumann entropy},\ }\href {https://doi.org/10.1088/1674-1056/19/4/040308}
	{\bibfield  {journal} {\bibinfo  {journal} {Chinese Physics B}\ }\textbf
		{\bibinfo {volume} {19}},\ \bibinfo {pages} {040308} (\bibinfo {year}
		{2010})}\BibitemShut {NoStop}%
	\bibitem [{\citenamefont {Cohen-Tannoudji}\ \emph {et~al.}(2019)\citenamefont
		{Cohen-Tannoudji}, \citenamefont {Diu},\ and\ \citenamefont
		{Lalo{\"e}}}]{Book_Cohen2019QuantumMechanicsVol1}%
	\BibitemOpen
	\bibfield  {author} {\bibinfo {author} {\bibfnamefont {C.}~\bibnamefont
			{Cohen-Tannoudji}}, \bibinfo {author} {\bibfnamefont {B.}~\bibnamefont
			{Diu}},\ and\ \bibinfo {author} {\bibfnamefont {F.}~\bibnamefont
			{Lalo{\"e}}},\ }\href
	{https://www.wiley.com/en-ie/Quantum+Mechanics,+Volume+1:+Basic+Concepts,+Tools,+and+Applications,+2nd+Edition-p-9783527345533}
	{\emph {\bibinfo {title} {Quantum Mechanics, Volume 1: Basic Concepts, Tools,
				and Applications}}}\ (\bibinfo  {publisher} {John Wiley \& Sons},\ \bibinfo
	{year} {2019})\BibitemShut {NoStop}%
	\bibitem [{\citenamefont {Hall}(2013)}]{Book_Hall2013Quantum}%
	\BibitemOpen
	\bibfield  {author} {\bibinfo {author} {\bibfnamefont {B.~C.}\ \bibnamefont
			{Hall}},\ }\href {https://link.springer.com/book/10.1007/978-1-4614-7116-5}
	{\emph {\bibinfo {title} {Quantum theory for mathematicians}}},\ Vol.\
	\bibinfo {volume} {267}\ (\bibinfo  {publisher} {Springer},\ \bibinfo {year}
	{2013})\BibitemShut {NoStop}%
\end{thebibliography}
